\documentclass[twocolappendix,dvipsnames,twocolumn]{aastex63}

\usepackage{graphicx}
\usepackage[abs]{overpic}

\usepackage{chngcntr}

\usepackage{amsmath}
\usepackage{gensymb}
\usepackage{booktabs} %
\usepackage{tabularx} %
\usepackage{array}

\usepackage[inline,shortlabels]{enumitem}

\usepackage{xspace}

\usepackage{rotating} %

\let\orgautoref\autoref

\renewcommand{\autoref}
{\def\equationautorefname{Eq.}%
\def\figureautorefname{Fig.}%
\def\subfigureautorefname{Fig.}%
\def\chapterautorefname{Ch.}%
\def\sectionautorefname{Sect.}%
\def\subsectionautorefname{Sect.}%
\def\subsubsectionautorefname{Sect.}%
\def\Itemautorefname{item}%
\def\tableautorefname{Table}%
\orgautoref}

\providecommand{\autorefs}
{\def\equationautorefname{Eqs.}%
\def\figureautorefname{Figs.}%
\def\subfigureautorefname{Figs.}%
\def\chapterautorefname{Chs.}%
\def\sectionautorefname{Sects.}%
\def\subsectionautorefname{Sects.}%
\def\subsubsectionautorefname{Sects.}%
\def\Itemautorefname{items}%
\def\tableautorefname{Tables}%
\orgautoref}

\newcommand{\ie}{i.e.\/,\xspace}

\newcommand{\eg}{e.g.\/,\xspace}

\newcommand{\Sim}{\sim\kern-0.2em\xspace}

\newcommand{\mrm}[1]{\ensuremath{\text{#1}}}

\newcommand{\dgr}{\ensuremath{\degree}\xspace}

\newcommand{\veritas}{{VERITAS}\xspace}
\newcommand{\fermi}{{\textit{Fermi}}\xspace}
\newcommand{\fermil}{{\textit{Fermi}-LAT}\xspace}

\newcommand{\cta}{{CTAO}\xspace}

\newcommand{\irf}{IRF\xspace}
\newcommand{\irfs}{IRFs\xspace}

\newcommand{\rnn}{RNN\xspace}
\newcommand{\rnns}{RNNs\xspace}

\newcommand{\mwl}{multiwavelength\xspace}

\newcommand{\offsetweightdecay}{\ensuremath{\gamma_{\mrm{decay}}}\xspace}
\newcommand{\offsetweightdelay}{\ensuremath{t_{\mrm{delay}}}\xspace}

\newcommand{\tref}{\ensuremath{t_{\mathrm{ref}}}\xspace}

\newcommand{\gammaerr}{\ensuremath{\gamma_{\mrm{err}}}\xspace}
\newcommand{\gammasignoise}{\ensuremath{\gamma_{\mrm{signoise}}}\xspace}
\newcommand{\gammatimeaggr}{\ensuremath{\gamma_{\mrm{timeaggr}}}\xspace}

\newcommand{\fref}{\ensuremath{F_\mathrm{ref}}\xspace}

\newcommand{\sstitlenoskip}[1]{\noindent\textbf{#1.\/}}

\newcommand{\framework}{the framework\xspace}

\newcommand{\encoder}{\ensuremath{\tau_{\textrm{enc}}}}
\newcommand{\decoder}{\ensuremath{\tau_{\textrm{dec}}}}
\newcommand{\forecast}{\ensuremath{\tau_{\textrm{for}}}}
\newcommand{\residuals}{\ensuremath{\tau_{\textrm{res}}}}
\newcommand{\weightedresiduals}{\ensuremath{\tau'_{\textrm{res}}}}
\newcommand{\reconstruction}{\ensuremath{\tau_{\textrm{rec}}}}
\newcommand{\temporalweights}{\ensuremath{\omega}}
\newcommand{\tsrec}{\ensuremath{\mathrm{TS}_\mathrm{rec}}}
\newcommand{\tsmm}{\ensuremath{\mathrm{TS}_\mathrm{mm}}}

\newcommand{\sigmarec}{\ensuremath{\sigma_\mathrm{rec}}}
\newcommand{\sigmamm}{\ensuremath{\sigma_\mathrm{mm}}}
\newcommand{\sigmacombo}{\ensuremath{\sigma_\mathrm{combo}}}

\begin{document}

\title{Early Detection of Multiwavelength Blazar Variability}

\correspondingauthor{Hermann Stolte (\href{mailto:hermann.stolte@hu-berlin.de}{hermann.stolte@hu-berlin.de}), Jonas Sinapius (\href{mailto:jonas.sinapius@desy.de}{jonas.sinapius@desy.de})}

\author[0000-0003-0047-5842]{Hermann Stolte}
\affiliation{Humboldt-Universität zu Berlin, Unter den Linden 6, 10117 Berlin, Germany}
\author[0009-0004-8608-0853]{Jonas Sinapius}
\affiliation{Deutsches Elektronen-Synchrotron DESY, Platanenallee 6, 15738 Zeuthen, Germany}
\author[0000-0003-1387-8915]{Iftach Sadeh}
\affiliation{Deutsches Elektronen-Synchrotron DESY, Platanenallee 6, 15738 Zeuthen, Germany}
\author[0000-0002-0529-1973]{Elisa Pueschel}
\affiliation{Ruhr-Universität Bochum, D-44780 Bochum, Germany}
\affiliation{Deutsches Elektronen-Synchrotron DESY, Platanenallee 6, 15738 Zeuthen, Germany}
\author[0000-0003-3325-7227]{Matthias Weidlich}
\affiliation{Humboldt-Universität zu Berlin, Unter den Linden 6, 10117 Berlin, Germany}
\author[0000-0002-2918-1824]{David Berge}
\affiliation{Humboldt-Universität zu Berlin, Unter den Linden 6, 10117 Berlin, Germany}
\affiliation{Deutsches Elektronen-Synchrotron DESY, Platanenallee 6, 15738 Zeuthen, Germany}

\begin{abstract}

Blazars are a subclass of active galactic nuclei with relativistic jets pointing toward the observer. They are notable for their flux variability at all observed wavelengths and timescales. Together with simultaneous measurements at lower energies, the very-high-energy (VHE) emission observed during blazar flares may be used to probe the population of accelerated particles. However, optimally triggering observations of blazar high states can be challenging. Notable examples include identifying a flaring episode in real time and predicting VHE flaring activity based on lower-energy observables. For this purpose, we have developed a novel deep learning analysis framework, based on data-driven anomaly detection techniques. It is capable of detecting various types of anomalies in real-world, multiwavelength light curves, ranging from clear high states to subtle correlations across bands. Based on unsupervised anomaly detection and clustering methods, we differentiate source variability from noisy background activity, without the need for a labeled training data set of flaring states. The framework incorporates measurement uncertainties and is robust given data quality challenges, such as varying cadences and observational gaps. We evaluate our approach using both historical data and simulations of blazar light curves in two energy bands, corresponding to sources observable with the Fermi Large Area Telescope and the upcoming Cherenkov Telescope Array Observatory. In a statistical analysis, we show that our framework can reliably detect known historical flares.

\end{abstract}


\section{Introduction}
Observations of transient phenomena are key to disentangling the physical processes at play in astrophysical systems. In the past years, notable transient events include the spatial correlation of an astrophysical neutrino with a flaring gamma-ray source~\citep{Aartsen:2018}, the observation of gamma rays and gravitational waves produced by a kilonova~\citep{PhysRevLett.119.161101}, and the discovery of new classes of very-high-energy (VHE; $>$100 GeV) gamma-ray emitters such as gamma-ray bursts and novae~\citep{MAGICGRB, 2019Natur.575..464A, Aharonian:2022} via detection of transient emission. Measuring or constraining the very-high-energy (VHE) gamma-ray emission from transient events is of particular interest, as these gamma rays track the most extreme acceleration processes.

Imaging atmospheric Cherenkov telescopes (IACTs), such as VERITAS, MAGIC, H.E.S.S., and the next generation Cherenkov Telescope Array Observatory, are the most sensitive instruments for measuring VHE gamma-ray emission~\citep{VERITASinstrument, MAGICinstrument, HESSinstrument, CTAinstrument}. However, they have fields of view of less than 10$\dgr$, which limits the chance of serendipitous transient detection. Source variability detected at other wavelengths is therefore used to trigger IACT observations, increasing the probability of observing transient events with IACTs. The deep learning approach presented here collates \mwl (MWL) information on the activity state of known gamma-ray sources and uses this information to anticipate periods of unusual emission (\eg flares) in the VHE band. In developing this method, we focus on blazars, a gamma-ray source class which shows strong variability on all observed timescales and wavelengths, and for which a large archive of VHE and MWL observations exists. The code used in this work is available on GitLab at \url{http://gitlab.desy.de/trans_finder/blazar_flares}.

\pagebreak

\subsection{Blazars}
Blazars are a class of active galactic nuclei with a relativistic jet oriented towards the observer, resulting in strongly Doppler-boosted emission. They are the most numerous source class detected in high energy (HE; 100 MeV--100 GeV) and VHE gamma rays. However, the mechanisms driving particle acceleration in the jets and the observed gamma-ray emission are not well-understood, and models of varying complexity abound (for discussion of acceleration see \eg \cite{Kirk_2000, Sironi_2015}; for emission see \cite{Cerruti_2020} and references therein). 

Gaining a better understanding of the underlying acceleration and emission mechanisms in blazars is relevant for a number of topics. For example, blazars are plausible sources of ultra-high-energy cosmic rays, should protons be accelerated in their jets (\eg \cite{Murase_2012, Rodrigues_2018}). As extragalactic sources, located at cosmological distances from Earth, blazar observations can be used to probe the photon and magnetic fields traversed en route to Earth, and to test for signatures of Lorentz invariance violation and photon coupling to axion-like particles, effects that are expected to grow with gamma-ray propagation distance (see \cite{galaxies10020039} and references therein). For such scientific goals, precise measurements of the broadband spectral energy distributions of sources (SEDs) is critical, being the main handle for interpretation of the emission mechanisms.

The photon emission from all VHE-detected blazars is characterized as variable at some or all observed wavelengths, from radio to VHE, on timescales from minutes to years. 
Bright states are of particular interest, but it is challenging to simultaneously detect a source and measure the SED over multiple wavelengths. 
This is due in part to limitations on collection areas of instruments in different wavelengths, as well as to the intrinsic variation in flux as a function of energy. 
Sources that are typically too dim to be detected in the HE and VHE bands without long integration periods (weeks to months) can conversely be detected during flaring episodes within minutes or hours. 

The SED of a variable blazar is modelled most robustly using simultaneous MWL observations, in order to ensure that different time-dependent flux states are not conflated. Bright states are key to probing the emission mechanism of blazars, and to their physical interpretation. Particularly important is the VHE band, where the photon flux is comparatively low.
Flare detection is therefore a cornerstone of the scientific programs of the IACT community (see \cite{Aharonian_2007, ALBERT2008253, Acciari_2009, Abeysekara_2015} for selected examples). 
Identifying correlated variation at different wavelengths is also valuable, as different models of blazar emission predict correlations (or lack thereof) between different bands.
However, consistent definition and identification of flares
remains a challenge in the community, as \eg discussed by \cite{2024MNRAS.534.2142Z}.

\subsection{Deep Learning and Anomaly Detection}
Machine learning and, in particular, deep learning, are widely used in astronomy.
Many applications are based on supervised learning.
This involves input data which
are labeled, where the algorithm is trained to
encode the mapping between inputs and labels.
Examples include classification, such as
$\gamma$/hadron separation for IACT experiments \citep[see \eg][]{Feng_Lin_2016_iact_ml, NietoCastano_2017_iact_ml},
as well as regression tasks, \eg the evaluation of
the redshift of a galaxy~\citep{Sadeh:2015lsa}.
Conversely, for unsupervised learning,
specific labels are not known a priori
and the objective is to find patterns in the data.
Applications are commonly based on clustering~\citep{min2018survey}
and/or outlier detection~\citep{2018MNRAS.476.2117R, 2021MNRAS.502.5147M}.

In the following, we expand on the work of \cite{2020ApJ...894L..25S}
(hereafter SA20), who
utilized a recurrent neural network (\rnn) for outlier detection. 
An \rnn is a form of 
a directed graph, representing a sequence of steps in time.
Outputs from each time step are fed as input to the next,
in addition to the respective temporal data. Such models are useful for
characterization of complex data on different timescales.

SA20 illustrated the use of their method for the detection of 
various types of astrophysical transients,
such as gamma-ray bursts and neutrino emission by candidate neutrino point sources.
The general concept is to use an \rnn to characterize
the background to a potential transient. This is done
by providing the network with existing/past data, 
taken before the emergence of the putative transient event.
For instance, in the case of the search for an 
astrophysical neutrino transient, the background
represents the continuously observed signals of atmospheric neutrinos.
The purpose of the \rnn is to predict the background within the near future
(upcoming time steps of the network),
based on observations of the near past.
Potential transients are then detectable as deviations of new
observations from the expected background predictions. 

For their chosen datasets,
SA20 could construct \rnn inputs that increase with 
the intensity of putative signals.
For example, they used the reconstructed
gamma-ray flux within a region of interest from an IACT experiment,
which increases when a new gamma-ray source appears.
As such, transients would always manifest themselves 
as upward fluctuations from the background
predictions of the network. Considering observables that by construction
always scale up with the strength of the signal enables the definition of
a simple outlier score,
acting as a test statistic (TS) for detection. The latter
is defined by SA20 as 
the integrated difference between the predictions
of the \rnn and the stream of real data.

In the current work, we focus on blazar flares as the
target transient phenomena. 
However, we set out to identify any unexpected activity from the source, such as notable downward deviations or correlated changes of the emission on different timescales.
For this purpose, the simple test statistic used by SA20
is not appropriate,
given that it is designed exclusively
for upward deviations from the background. 
In order to construct a generalized test statistic for
arbitrary types of flares, we expand 
the architecture of the network used by SA20, as discussed in the following.

This paper is organized as follows.
In \autoref{secpipeline}, we introduce our machine-learning based framework and in \autoref{subsec:training} we describe how it is trained. \autoref{sec:simulation} contains a simulation study, in which we evaluate the sensitivity of our method. In \autoref{sec:realdata} we demonstrate the utility of our method in an analysis of real data from BL Lacertae, before concluding in \autoref{sec:discussion}.

\section{Anomaly detection pipeline}
\label{secpipeline}

\begin{figure*}[t]
    \centering
    \includegraphics[width = 0.95\textwidth]{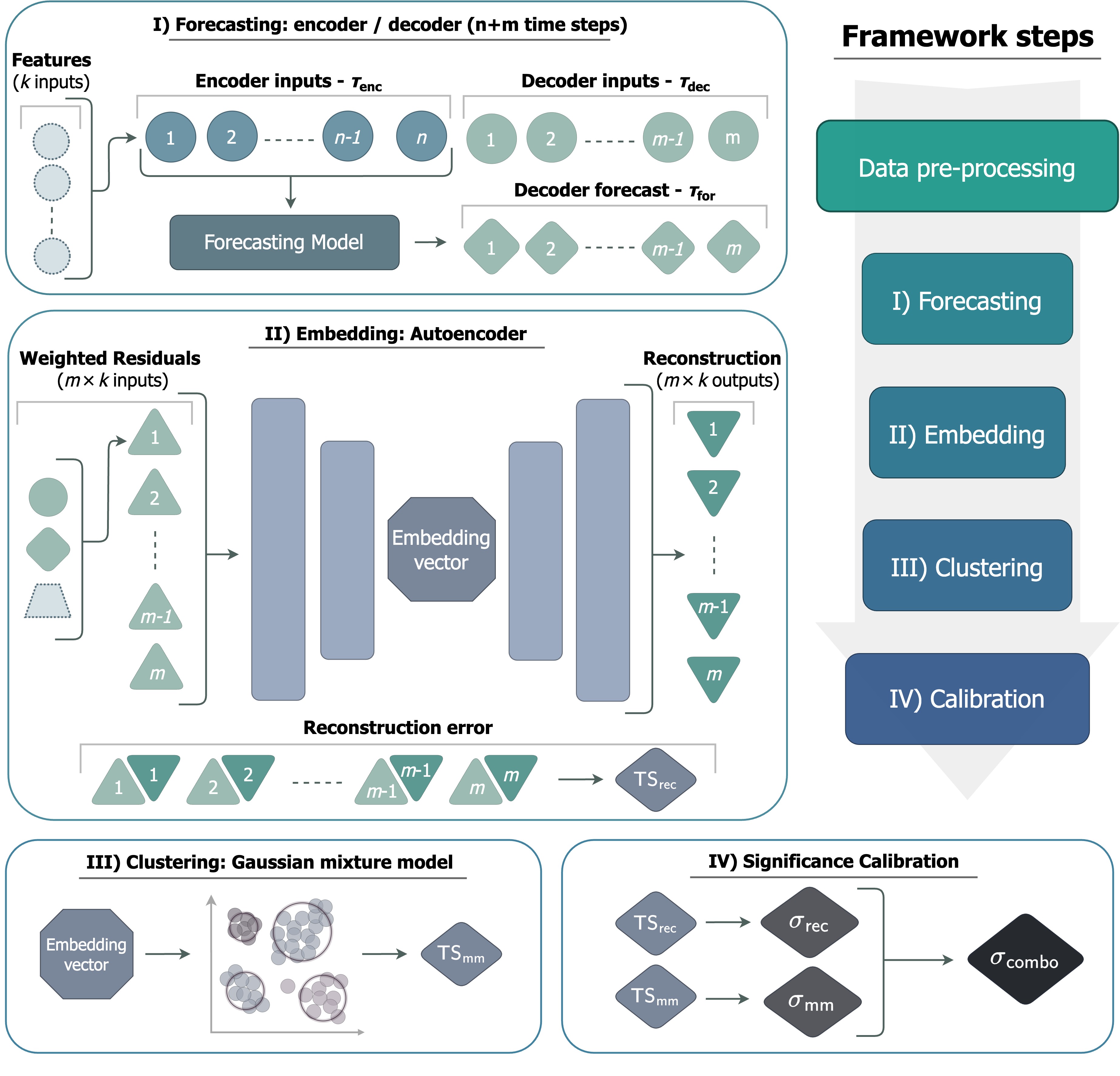}
    \caption{Illustration of the architecture of the model, as described in the text.  }
    \label{fig:model_arch}
\end{figure*}

Our pipeline is illustrated in \autoref{fig:model_arch} and 
the most important concepts and parameters are summarized in \autoref{tab:notations}. 
After preprocessing of the input light curves, 
the data are organized into
a time-series format. 
We distinguish between a \textit{context window}
and a \textit{search window} for the analysis. 
In the latter, we scan for putative flares. 

\sstitlenoskip{Forecasting (Box~I in \autoref{fig:model_arch})}
The inputs within the context window, denoted
by $\tau_{\mathrm{enc}}$, and the search window, 
$\tau_{\mathrm{dec}}$, are respectively mapped into the
encoder and decoder components of an RNN, similar to the
approach of SA20.
The purpose of the encoder is to provide
contextual information (recent source activity) for the decoder.
We train the model under
the assumption that
the context window is devoid of potential signals, and
thus represents a realistic background for flares.
The outputs of the decoder, $\tau_{\mathrm{for}}$, are
the model predictions for the light curve, given
this background hypothesis.

\begin{table}[t]
	\caption{Overview of notions and parameters.}
	\label{tab:notations}
	\centering
	\vspace{-0.75em}
    \footnotesize
        \begin{tabular}{l @{\hspace{0.6em}} p{6.5cm}}
		\toprule
		Notation & Description\\ 
		\midrule
            \tref & Reference time pivot in the input light curve \\
            
		$\encoder, \, \decoder$ & Inputs split into encoder (context) and decoder (search) windows, comprising $n$ and $m$ time steps\\
  
            \forecast & Forecast of background activity for $m$ time steps\\
            
            $\residuals, \, \weightedresiduals$ & Residuals between decoder, \decoder, and forecast, \forecast; tagged variable  is adjusted by temporal weights\\
            
            \temporalweights & Temporal weights applied to decoder residuals\\
            
            \reconstruction & Reconstruction of the weighted residuals, \weightedresiduals \\
            
            $\phi$ & Embedding vector representing residuals,  \weightedresiduals \\
            
            $\tsrec, \, \sigmarec$& Test statistic (TS) and significance, based on the reconstruction error\\
            
            $\tsmm, \, \sigmamm$& TS and significance, based on the clustering model compatibility metric \\
            
            $\sigmacombo$ & Combined significance from \tsmm~ and \tsrec \\
            \gammatimeaggr & Time bin width (in days for the current study) to be considered a single time step\\
            \gammasignoise & Threshold (in units of standard deviation) to con-strain variability; used to quantify background states for training\\
            \offsetweightdecay & Temporal decay rate, used for suppressing old anomalous data\\
		\bottomrule

    	\end{tabular}
	\vspace{-1em}
\end{table}

\sstitlenoskip{Embedding (Box II)}
The predictions are contrasted with the actual data, 
in order to
quantify anomalies. 
As explained in \autoref{subsec:representation}, we derive the difference
between the two sets as 
$\tau_{\mathrm{res}} = \tau_{\mathrm{dec}} - \tau_{\mathrm{for}}$, emphasizing the most recent data in time through a weighting scheme. 
We model the distribution of $\tau_{\mathrm{res}}$ for the background using an autoencoder, which compresses the data into a low-dimensional space.
For the subset of data which comprises background configurations, $\tau_{\mathrm{res}}$ represents the intrinsic variations of the light curve in the absence of flares. Otherwise, $\tau_{\mathrm{res}}$ quantifies the strength of a potential anomaly.

\sstitlenoskip{Clustering (Box~III)}
The condensed representation provided by the
autoencoder
facilitates comparison between new data and
the reference distribution, which comprises all known
background states.
We parameterize the distribution of $\tau_{\mathrm{res}}$
for the background dataset with
a Gaussian mixture model (a form of clustering),
as discussed in \autoref{subsec:clustering}. 

\sstitlenoskip{Significance calibration (Box~IV)}
The compatibility of new data with the modeled
distribution of $\tau_{\mathrm{res}}$
is used for anomaly detection. We follow the approach
of SA20 in order to map the corresponding test statistic into a significance
for detection. A detailed description and derivation of this calibration is given in the Appendix of SA20. In addition, we derive another complementary test statistic
directly from the outputs of the autoencoder, as
discussed below. The final \textit{p}-value for flare detection
is derived from the combination of these two statistics (\autoref{subsec:calibration}).

Once a significant flare is detected, it raises 
the question of which elements in the data are driving the detection. 
We analyze the contribution of different data to the overall significance within the pipeline, as described in \autoref{def:channelcontributions}.
Below, the various elements of the pipeline are discussed in detail. 

\subsection{Light Curves as Input Data}
\label{subsec:pipelineinputs}

The \mwl inputs to our pipeline are a set of light curves, obtained from different instruments. Each  input (which we refer to in the following as \textit{channel}) captures an observable related to source brightness (\eg flux, magnitude), with associated
uncertainties.

A potential flare is always searched for within a specific temporal context.
The data collection process behind each channel, however, is typically affected by numerous observational constraints, such as the visibility of the source and weather constraints for ground-based instruments. Consequently, the observing cadences commonly vary across channels and over time. Frequently, limited data availability and long gaps in coverage need to be handled. 
To address these temporal dynamics, we employ a series of techniques, which we discuss in the following.

\sstitlenoskip{Consistent time binning}
To jointly analyze observables in the timescales that are relevant for blazar flares, we first ensure a consistent time binning in each channel. Specifically,  we contrast between high-density and low-density channels, given the selected width of our time-binning, \gammatimeaggr (\eg one day,  customizable per channel). High-density channels, which generally
contain multiple data points within \gammatimeaggr, are downsampled by randomly selecting a single data point per \gammatimeaggr. This effectively captures the average state in the channel.

\sstitlenoskip{Dynamic windowing}
A potential flare is searched for in the data that is available up to a time pivot, \tref. 
Observation cadences and the corresponding coverage will generally vary across inputs and over time. The number of data points within a fixed time window may therefore differ between channels. 
To effectively interpret source activity in context, it may be required to consider older data from sparse channels together with recent data from dense channels. 
Instead of relying on fixed duration windows, it is hence desirable to consider time windows of variable lengths.  
We employ a dynamic windowing approach; it considers a fixed number of the most recent available data from each channel, up to \tref. 
The first (\ie oldest) $n$ data points in the window constitute the context window for flare detection, $\tau_{\textrm{enc}}$; the last (\ie newest) $m$ data points form the search window, $\tau_{\textrm{dec}}$. 

\sstitlenoskip{Variable and Sparse  Coverage}
In order to constrain emission models and predict future flares, 
the observation time of an individual measurement is crucial. 
However, the provenance of the available \mwl data
is subject to potentially arbitrary constraints. 
The cadence of measurements is influenced by several confounding factors, including observing conditions; source visibility; 
and instrumental availability.
Correspondingly, the presence of a particular measurement 
could itself be seen as abnormal.
While our framework is designed to consider any type of anomaly in the input channels, it must be agnostic to changes in data cadence, while modeling the relevant temporal context. 
For this reason, we consider the difference in cadence between measurements only indirectly through a dynamic weighting scheme, as described in the following.

In the assessment of potential flaring activity in channels with varying sparsity, recent data are generally considered more relevant than older data. 
This is particularly important for low-density channels, which generally contain a single data-point within a period of several \gammatimeaggr. 
The search window of a low-density channel may consist entirely of data that lie far in the past (relative to \tref). It may thus capture flaring behavior that is irrelevant for assessing the current activity of the source.   

We use a
scheme of temporal 
weights to
ensure the sensitivity to potential new flares, given the possible presence of old
anomalies. 
Weights are
defined as
\begin{equation}
    \omega(t_{\mathrm{step}}, \offsetweightdecay) = ( 1 - \mathrm{max}(0, t_{\mathrm{step}} - \offsetweightdelay))^{-\offsetweightdecay}.
\end{equation}

Here $t_{\mathrm{step}}$ is the temporal distance of a particular data point relative to $\tref$.
We define \offsetweightdelay as the period during which all data are considered recent. This parameter is defined by the number of decoder steps, $m$, and the time step interval, \gammatimeaggr, as
\begin{equation}
\offsetweightdelay = m \times \gammatimeaggr.
\end{equation}
The parameter, $\offsetweightdecay$, controls the strength of the temporal decay.

The temporal weights are used to scale the inputs to the autoencoder.
Weights are constructed, such that they are equal to unity in search windows that are not sparse.
This scheme serves to relatively shift the
residuals for older data towards background-level values. 
Thus, 
the impact of anomalies which are driven by source-variability in the far past is suppressed.

\begin{figure}
    \centering
    \includegraphics[width = 1.\columnwidth]{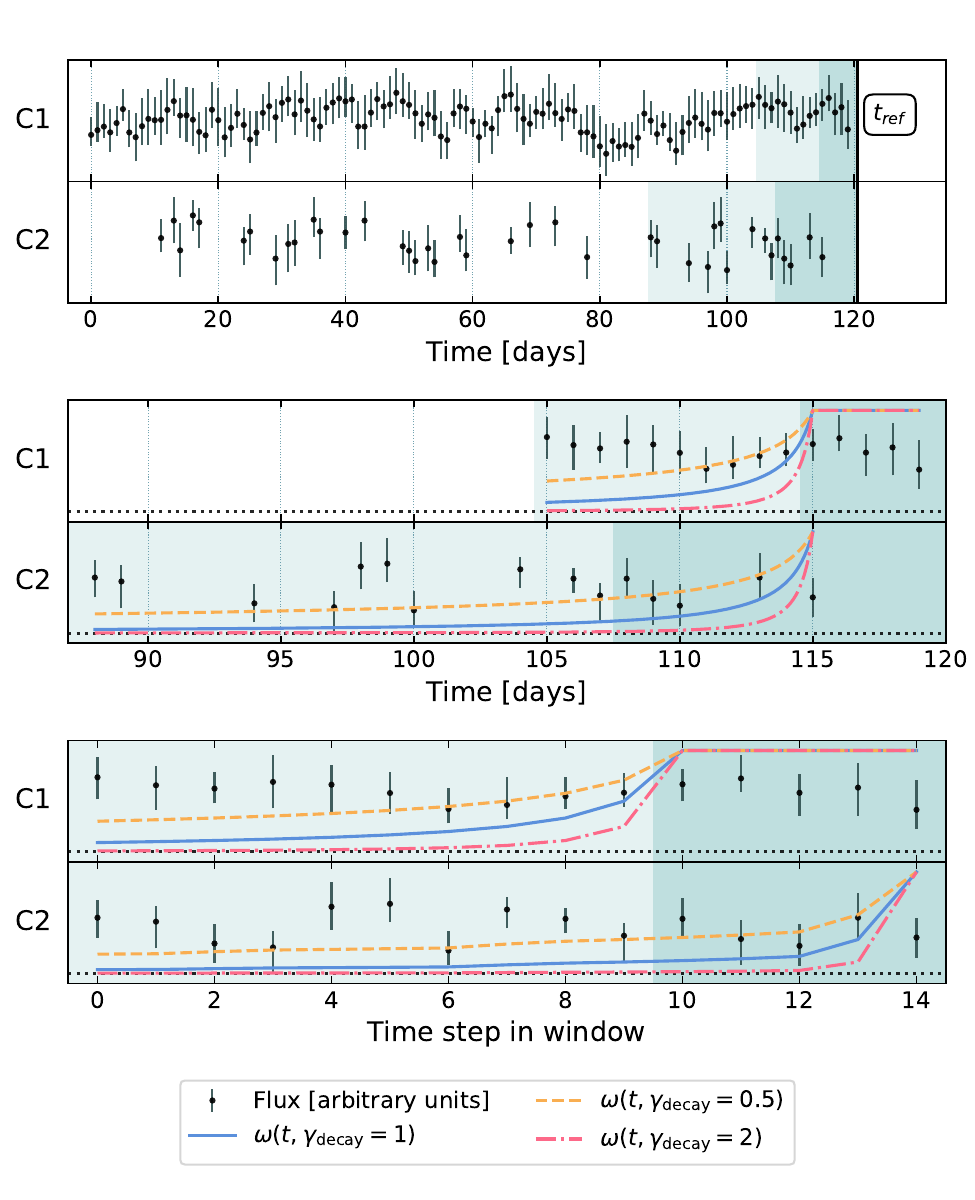}
    \caption{Illustration of a dynamic-window transformation of two light curves into a combined time series structure. 
    The top panel shows two light curve channels, C1 and C2, with a reference time pivot, \tref. The context window, $\tau_{\mathrm{enc}}$, and the search  window, $\tau_{\mathrm{dec}}$, are highlighted as light and dark shaded regions. 
    The middle panel shows a zoom in of both windows over the time relative to \tref. 
    The bottom panel shows both windows in $n+m$ discrete time steps, as they are ingested into the forecaster and autoencoder. 
    Colored lines in the lower panels show the corresponding temporal weights, $\omega$, for three representative settings of \offsetweightdecay. 
    The dotted horizontal line marks a weight, $\omega$, of zero. 
    }
    \label{fig:model_input}
\end{figure}

The temporal weighting scheme is illustrated in \autoref{fig:model_input}.
Here, the top panel shows an example input light curve comprising two channels and a time pivot, $\tref$. 
The first channel has dense coverage and thus a short window size. 
The second channel is more sparsely sampled. The corresponding window covers a wider time interval, as required to capture $n+m$ data points. 
The lower panel shows the corresponding time-series structure separated into the context window, $\tau_{enc}$, and the search window, $\tau_{dec}$. 
The dynamic weighting scheme and the effect of $\offsetweightdecay$ is also illustrated in the two bottom panels. 
Here, the data in the sparse channel are shifted towards background within the decoder, as there exists a notable gap between the most recent data and $\tref$. 
In contrast, the weights of the dense channel leave the data within the decoder unmodified.

\sstitlenoskip{Limited data availability}
The availability of data in a channel may be limited, for instance due to novel instruments starting operations, or to 
having made only few observations of a source. 
In some cases, a channel may contain less data than required for the construction of a single window. To ensure the compatibility of our approach with limited data, we add padding to the beginning of the window. It consists of new data points, derived from existing data, randomly scrambled in time.
This includes timestamps lying in the far past. With the aforementioned weighting scheme, we ensure that the padding segment of data does not result in anomalies.

\sstitlenoskip{Uncertainty on observables}
In order to account for the uncertainties on individual measurements, we draw a sample of \gammaerr time series
with feature values sampled from the assumed normally distributed, potentially asymmetric uncertainties of the data. While this is a common approach, the framework remains agnostic to the choice of distribution used for modeling uncertainties on observables. The following stages independently process the original time series as well as all uncertainty samples. In the final calibration, as discussed below, their corresponding test statistics are aggregated. The final outcome is then a calibrated significance with an associated uncertainty, which represents the propagated input uncertainties from all channels.

\subsection{Forecasting Background Activity}
\label{subsec:forecasting}

After preprocessing, the input light curves are transformed into fixed-size windows that follow a time series structure of
$n+m$ time steps. Each step 
contains $k$ features, one for each input channel. The specific values chosen for these parameters are specified in \autoref{simtraining}. As illustrated in  \autoref{fig:model_arch} - Box I, a window is then split into two parts, an encoder, $\tau_{\mathrm{enc}}$, and a decoder, $\tau_{\mathrm{dec}}$. 
The encoder serves as the reference for a forecast, $\tau_{\mathrm{for}}$, representing the predicted background-level source activity. Both $\tau_{\mathrm{dec}}$ and $\tau_{\mathrm{for}}$ comprise the same $m$ time steps following the $n$ time steps in $\tau_{\mathrm{enc}}$. 
To obtain the forecast of background activity, we employ a multivariate, multistep \rnn with probabilistic output layers, similar to the encoder-decoder architecture utilized by SA20. 

\subsection{Representation Learning}
\label{subsec:representation}

After the forecasting of background activity, we
contrast the actual data with the background forecast, $\tau_{\mathrm{for}}$. That is, we derive element-wise residuals  as 
\begin{equation}
    \tau_{\mathrm{res}} = \tau_{\mathrm{dec}} - \tau_{\mathrm{for}}.
\end{equation}
Each element in $\residuals$ is then multiplied with its corresponding temporal
weight, 
\begin{equation}
    \tau_{\mathrm{res}}' =
    \temporalweights \cdot \tau_{\mathrm{res}},
\end{equation}
where the elements of $\tau_{\mathrm{res}}'$ serve as the inputs to the
next element of the model.

This definition of $\tau_{\mathrm{res}}'$ is motivated
by the fact that small numerical values of the residuals
correspond to background states, 
while larger positive or negative values represent increasingly
significant anomalies. 
Effectively, this weighting scheme shifts older data towards background states, with $\omega=1$ corresponding to the original data and $\omega=0$ being fully compatible with background.

\color{black}
Based on the weighted residuals, $\tau_{\mathrm{res}}'$, we obtain an embedding vector, $\phi$, that represents the source variability across channels. 
The purpose of the mapping into an embedding vector is to capture a condensed representation of variability across all channels. 
It enables the comparison of  relative distances between source activity states, which is fundamental to one of our two test statistics, $\mathrm{TS}_\mathrm{mm}$, used for flare detection (see \autoref{subsec:clustering}). 

To map $\tau_{\mathrm{res}}'$ into an embedding vector, we employ a variational autoencoder with probabilistic output layers, based on \rnns. 
As the autoencoder is a modeled distribution of observed source activity, it can fail to generalize to novel source activity.
This could happen, if the novel activity is too different from the data used for training, and would result in unreliable mappings into the embedding space. 
We interpret such a case as a potential type of flaring state, where the differences to known source activity are so pronounced, that a comparison in the embedding space is infeasible. 

To reliably detect this type of flare, we  compare the reconstruction outputs from the autoencoder, $\tau_\mathrm{rec}$, to the inputs, $\tau_{\mathrm{res}}'$. 
When the embedding vector is unreliable, the corresponding reconstruction error is high. 
Hence, we define a test statistic, $\mathrm{TS}_\mathrm{rec}$, as the compatibility of each output with its corresponding input distribution. 
Specifically, this test statistic is calculated as the summed negative log likelihood of each element in $\tau_{\mathrm{res}}'$ with respect to the corresponding distribution reconstructed from the embedding vector, 
\begin{equation}
    \mathrm{TS}_\mathrm{rec}(\tau_{\mathrm{res}}', \tau_{\mathrm{rec}}) = \sum_{i=0}^{m} \sum_{j=0}^{k} -\mathrm{log} \ p(\tau_{\mathrm{res}_{i,j}}' | \tau_{\mathrm{rec}_{i,j}})
\end{equation}
The mapping into embedding space and derivation of $\mathrm{TS}_\mathrm{rec}$ is illustrated in \autoref{fig:model_arch} (Box~II). 

\color{black}
\subsection{Clustering}
\label{subsec:clustering}

The embedding vector, $\phi$, obtained from the autoencoder, represents an unknown state of source activity. Next, we calculate a test statistic, $\mathrm{TS}_\mathrm{mm}$, that quantifies the compatibility of this state with a model for the background. 
We use a Bayesian Gaussian mixture model 
\citep[BGMM;][]{mclachlan2000finite,10.1214/06-BA104}{}{}%
, $\chi$, 
for modeling background activity in the embedding space. 
The goal is to allow for complex clusters of different background states to be jointly modeled.  

The calculation is illustrated in \autoref{fig:model_arch} - Box III. Intuitively, one can relate $\mathrm{TS}_\mathrm{mm}$ to the geometric distance between the embedding vector of a source state and the distribution of modeled background states derived from the BGMM. The ``farther away'' the source state embedding lies from the background embeddings, the less likely it is for this source state to be compatible with the background model.
Technically, the test statistic is derived as the negative log probability of the embedding vector with respect to the mixture model,

\begin{equation}
    \mathrm{TS}_\mathrm{mm}(\phi) = -\mathrm{log} \ p(\chi | \phi)
\end{equation}

\subsection{Significance Derivation}
\label{subsec:calibration}

Box IV in \autoref{fig:model_arch}, illustrates the final step
of the pipeline, carried out after the
model has been fully trained. 
Here we map the two test statistics,  
$\mathrm{TS}_\mathrm{mm}$ 
and $\mathrm{TS}_\mathrm{rec}$, 
into their respective \textit{p}-values for anomaly detection,
$p^{\mathrm{v}}_\mathrm{mm}$ and $p^{\mathrm{v}}_\mathrm{rec}$. 
As defined above,
the first metric, $\mathrm{TS}_\mathrm{mm}$,
represents in-distribution anomalies (low compatibility with the background model).
The second metric,
$\mathrm{TS}_\mathrm{rec}$,
captures out-of-distribution anomalies (large
reconstruction errors of
the autoencoder).
The \textit{p}-values are conveniently 
represented by their
corresponding significance values,
the clustering significance, $\sigma_\mathrm{mm}$, and the reconstruction significance,
$\sigma_\mathrm{rec}$. 

In addition to these two
quantities, we derive
a high-level test statistic. 
This metric is based
on the two individual \textit{p}-values, 
and is defined as
\begin{equation}
    \mathrm{TS}_\mathrm{comb}
        = -\mathrm{log} \ p^{\mathrm{v}}_\mathrm{rec}
        -\mathrm{log} \ p^{\mathrm{v}}_\mathrm{mm} \; .
\end{equation}
The combined test statistic is similarly used to
derive the final \textit{p}-value and
significance for the pipeline, 
$p^{\mathrm{v}}_\mathrm{combo}$
and $\sigma_\mathrm{comb}$.

The so-called \textit{calibration} process between each test statistic
and the related significance for detection 
follows the prescription of SA20.
We employ a numerical approach
that involves evaluating each test statistic
multiple times, given a representative sample
of background examples. 
The corresponding
distribution of test statistics therefore
represents the 
null-hypothesis for anomalies. 
The test statistics are constructed such that high numerical values
correspond to larger potential anomalies.
The distribution for the
background dataset may therefore be used to estimate
the probability for a high TS value
(small \textit{p}-value), and the related significance.

\subsection{Characterization of Signals}
\label{def:channelcontributions}
When a flare is identified by \framework, retracing what aspects of the data contribute to this identification is nontrivial. However, understanding what part of the overall behavior is considered anomalous is key in furthering our understanding of the nature of blazar flares. 
An anomaly in the behavior of a single channel can constitute a flare, as well as the combined behavior of multiple channels. For instance, a correlation between two channels may be anomalous, even in cases where the same data would not individually be flagged as unusual.  

To guide the identification of anomalous behavior, we assess the significance when considering only subsets or individual channels. This is accomplished by modifying the temporal weights, $\omega$, on $\weightedresiduals$. Specifically, we set $\omega=0$ for all channels that are not considered in a given iteration during inference. This is effectively a boolean mask, that sets all data from masked channels to correspond to background. In the evaluation, we demonstrate how this enables the tracing of significant detections back to individual input channels. 

We note that these projections in the embedding space are not individually calibrated to \textit{p}-values. That is, the corresponding significance of individual channels is indicative of their relative contribution to anomalies. However, they should not be interpreted as the final significance for detection.

\section{Training the model}
\label{subsec:training}
All components of the pipeline are trained on data corresponding to background source activity. The available data, however, typically consist of light curves which include flaring periods or other anomalies. 
To obtain a background-only version of the data, we rely on a combination of (optional) manual filtering to remove obvious flaring states, as well as several automated cleaning and augmentation procedures. Both methods
are described in detail in   
\autoref{cleaninganomalies}.
They include several randomization procedures for statistically removing potential remaining anomalies from the data used for training (forecasting and clustering stages) as
well as for calibration. 
At this stage, 
the training dataset is normalized, such that
each channel (\ie feature) has
a mean of zero and a standard deviation of one.

The dataset is augmented in order to increase the
robustness of our model.
As described in  \autoref{longtermtrends}, we introduce randomly generated long-term states to a subset of
training examples.
As part of training the autoencoder, we also add randomly generated short-term fluctuations, as described in \autoref{fluctuations}. This augmentation ensures that the embedding space covers a wide range of source activity. It helps to reduce
potential reconstruction errors and improves the generalizability of the pipeline. 
By construction, these fluctuations deviate from the background distribution we intend to model.
We therefore emphasize that they are not used for fitting the BGMM or for the significance calibration process.

The training process proceeds as follows.

\sstitlenoskip{Step I} The forecasting model
is trained on background-only data using the Adam optimizer~\citep[][]{kingma2014adam}{}{} with the summed log probability for the loss; a learning rate of $0.005$; a dropout of $10\%$; L2 regularization of $0.0001$; and early stopping.

\sstitlenoskip{Step II} The autoencoder 
is trained analogous to the forecasting model. 
As a notable difference, the training dataset is enriched with injected short-term fluctuations. 
After training the forecasting model, the inputs to the autoencoder, \weightedresiduals,  are normalized to have a mean of zero and a standard deviation of one. 
The temporal weight decay, $\offsetweightdecay$, is set to randomized values within the range $\{1, 2\}$ to further enhance the phase space of the embeddings. 
The optimizer and regularization  techniques are analogous to the training of the forecasting stage. 

\sstitlenoskip{Step III} The Gaussian mixture model of background-level activity 
is fitted using a Markov chain Monte Carlo (MCMC) approach. Here, we adaptively increase the complexity of both the model and the MCMC parameters, until there is no notable improvement in the quality of fit. The dynamic parameters include the number of model components as well as the number of MCMC steps, of burning steps, and of step-size adaptations. The quality of fit is determined through a two-sample Kolmogorov-Smirnov test for equal distributions between the input and the posterior sample distribution. To further improve the convergence stability, multiple MCMC chains are fitted in parallel. The best chain is selected according to the quality of fit and is used for subsequent inference. Through adaptively increasing the complexity of the model and the fit, we ensure convergence once the process terminates regularly, \ie without reaching a timeout.   

\subsection{Cleaning Anomalies from the Training Dataset}
\label{cleaninganomalies}
Available datasets for training typically comprise light curves that include flaring periods or other anomalies. As the input to the training stage of the pipeline should consist only of background examples, such data are cleaned prior to training.
Well-defined and extreme flaring intervals are removed by hand. This procedure is optional. 

Following manual cleaning, some anomalies may remain in the data, such as weak, short-term fluctuations, or other correlations between channels. To ensure sensitivity to these types of anomalies, the pipeline includes
additional statistical randomization and cleaning steps. 

As part of automated cleaning, 
we use a variability threshold (in units of standard deviation), defined by the parameter, \gammasignoise.
When a particular window contains data with a local signal-to-noise ratio higher than \gammasignoise, we exclude it from the training and calibration stages. 
This ensures that the model becomes sensitive only to deviations stronger than the configured threshold. Tuning this parameter is crucial;  increasing the value of \gammasignoise reduces the rate of false positives, but also decreases the sensitivity to faint flares. On the other hand, lowering the threshold raises the rate of false positives, while improving the sensitivity to faint flares.

We also shuffle data
inside each window.
This suppresses residual correlated anomalies within the training dataset. 
It is important to
avoid introducing a bias to the pipeline by always having the same observation cadence in any one channel. We therefore randomize the timestamps of individual data points as follows.
Considering data points i and j taken at times $t_\mathrm{i} < t_\mathrm{j}$, we reassign the time\-stamp $t_\mathrm{i} \rightarrow t'_\mathrm{i}$ to a random value that satisfies the condition, $t_\mathrm{i} \leq t'_\mathrm{i} < t_\mathrm{j}$,  and conforms to the selected time bin width, \gammatimeaggr.  This is done for each data point in each input channel independently. 

Finally, we apply a randomized temporal shift to each light curve individually; this is done globally for the entire channel. Multiple randomized realizations of the dataset cover a wide range of relative global shifts between channels. Such transformations suppress possible inter-band correlations.

\subsection{Timescales for Anomaly Detection}
\label{longtermtrends}
Blazars are known to exhibit variability on short (minutes), intermediate (days), and long timescales (years). The examples featured in the current study are focused on detecting medium timescale behavior, on the order of a few days. 
We account for very short flares with the choice of $\gammatimeaggr = 1$~day.
It therefore remains important to consider the impact of long-term variability. 
Over long enough time periods, on the order of a few years, blazar emission can exhibit trends and correlated behavior of different levels of intensity. We would like to avoid false-positive detections for a medium-timescale search. Such slow-changing trends should therefore be made indistinguishable from background in the training dataset. 

We augment the training sample accordingly. 
We introduce randomly generated long-term states in the input light curves. This is done by globally shifting the data up or down by up to six standard deviations for a particular realization. Here, all data in a given channel have the same vertical offset applied, where individual channels are treated independently. This augmentation also serves to improve the generalization of the modeled background.

\subsection{Embedding Space Augmentation}
\label{fluctuations}
To allow the embedding space to represent different types and intensities of potential flares, we enrich the dataset for training the autoencoder by introducing arbitrary short-term fluctuations. 
These allow the embedding space to encode distances between different types of potential flaring states. 
Specifically, the random fluctuations are modeled as step functions and Gaussians with amplitudes randomly sampled within $\{2,100\}$ (upward) or $\{-2,-10\}$ (downward) local standard deviations. 
Their duration is within $\{1,8\}$ days. The temporal onset of a fluctuation is randomly shifted across channels by up to five days.

section{Simulation study}
\label{sec:simulation}
We evaluate the performance of our pipeline based on simulations of quiescent and flaring states inspired by the TeV-detected BL Lac object 1ES$~$1215$+$303.
We simulate and evaluate different flare scenarios, varying in duration and flux. 
The light curves cover observations from  \fermil and \cta. 

The Large Area Telescope (LAT) is a pair-conversion telescope on board the \fermi spacecraft \citep[][]{atwood_2009_fermi_lat}, covering an energy range from $\sim 20~$MeV to more than $500~$GeV. During standard operations, it covers the full sky once every three hours.  

The upcoming Cherenkov Telescope Array Observatory (\cta) is a collection of VHE gamma-ray IACTs, with one array on the southern hemisphere and one on the northern hemisphere \citep[][]{ctao_concept}. Covering energies between $\sim20~$GeV and $300~$TeV. Having  increased sensitivity compared to current IACTs, it will significantly improve knowledge of the VHE sky.

In \autoref{simdataset} we describe the simulation in more detail, before discussing the training process and pipeline parameters in \autoref{simtraining}. The results are shown in \autoref{simresults}.

\subsection{Dataset}
\label{simdataset}

\begin{figure}
    \centering
    \includegraphics[width=0.5\textwidth]{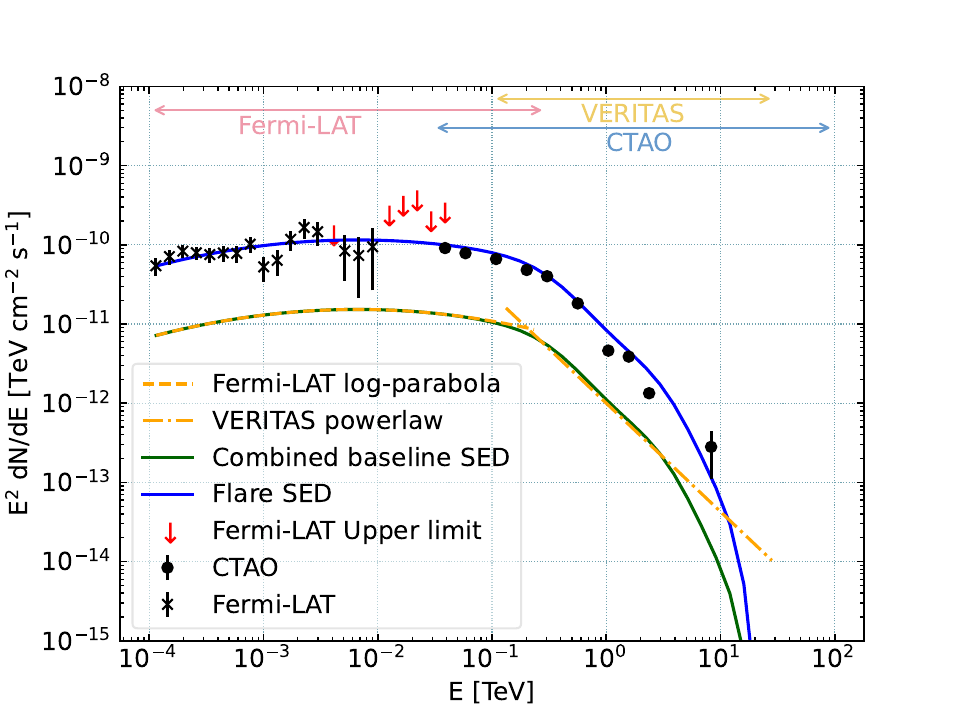}
    \caption{Spectral energy distributions for the simulated dataset, inspired by the 2017 steady-state spectrum of the blazar, 1ES$~$1215$+$303. Horizontal arrows show the sensitivity regions of \fermil, VERITAS, and \cta respectively. The orange dashed (dashed-dotted) line shows the spectrum given in \citet{Valverde_2020} for LAT (VERITAS), and the solid green (blue) line shows the combined, EBL-absorbed log-parabola used for the baseline (flaring) simulations.}
    \label{fig:sim_seds}
\end{figure}

We simulate the multi-band gamma-ray light curve of a quiescent source, mimicking the 2017 steady-state spectrum of 1ES$~$1215$+$303, as measured by \citet{Valverde_2020} with \fermil and VERITAS. The measured \fermil log-parabola and the VERITAS power law are jointly well-described with a log-parabola that accounts for extragalactic background light (EBL) absorption, shown in \autoref{fig:sim_seds}. 
Assuming constant source behavior, we use the \textit{gtobssim} tool from Fermitools\footnote{\url{https://github.com/fermi-lat/Fermitools-conda/}}, version 2.2.0, and the \textit{gammapy} package \citep[v 0.20.1,][]{gammapy:2017}{}{} to produce simulations of 1000-day daily-binned \fermil and \cta light curves, respectively. To simulate a realistic IACT observing cadence, we conservatively decrease the sampling of the \cta light curve to mimic the sampling of real VERITAS observations of this source. Subdividing the light curves into multiple, non-overlapping energy bins allows us to keep some of the spectral information. The \fermi band is divided into three energy bins 
(100 -- 669~MeV, 
669~MeV -- 4.48~GeV, and 
4.48 -- 300~GeV);
the \cta band is divided into four energy bins 
(31.6 -- 79.4~GeV, 
79.4 -- 199.5~GeV, 
199.5~GeV -- 1.2598~TeV, and
1.2598 -- 12.589~TeV). 
The inputs and specifics of these simulations are described in more detail in \autoref{App:SIM_DATA}.

We create templates to evaluate the performance on simulated flares, as described above. As a proxy for a high state, we use the 2017 flare studied by \citet{Valverde_2020}{}{}, denoted as $F_\mathrm{ref}$ in the following. For comparison, the SEDs of the baseline and the unscaled flare are shown in \autoref{fig:sim_seds}. We simulate observations of a flare with a top-hat profile on six subsequent days with 1\%, 10\%, 50\%, 100\% and 150\% of  $F_\mathrm{ref}$. By applying these templates only partially, we can also simulate flares with durations between one and five days. To increase the richness of the dataset and to better account for random fluctuations in these templates, we create ten randomized realizations for each template at each intensity.
While we are generally interested in sudden high-flux states in blazars, an unexpected decrease in flux can also be anomalous and  potentially interesting. To test how \framework reacts to such cases, we create realizations where we subtract these templates from the baseline.

\subsection{Configuration Parameters and Training}
\label{simtraining}

We utilize a time bin width of ${\gammatimeaggr = 1}$ day to transform the data into windows consisting of  ${m = 10}$ context steps and  ${n = 5}$ search steps. To filter non-background data, a signal-to-noise threshold of ${\gammasignoise = 5}$ is applied.
We generate 100 augmented realizations of pure background, and 100 augmented realizations including short-term fluctuations.  
The temporal weight decay is chosen as ${\offsetweightdecay = 1}$.
The training converges, which we confirm by inspecting the loss curves and calibration control plots for the goodness of fit. We also require that the adaptive MCMC fit terminates regularly. 

\subsection{Results}
\label{simresults}

\begin{figure*}[t]
    \centering
    \includegraphics[width=0.99\textwidth]{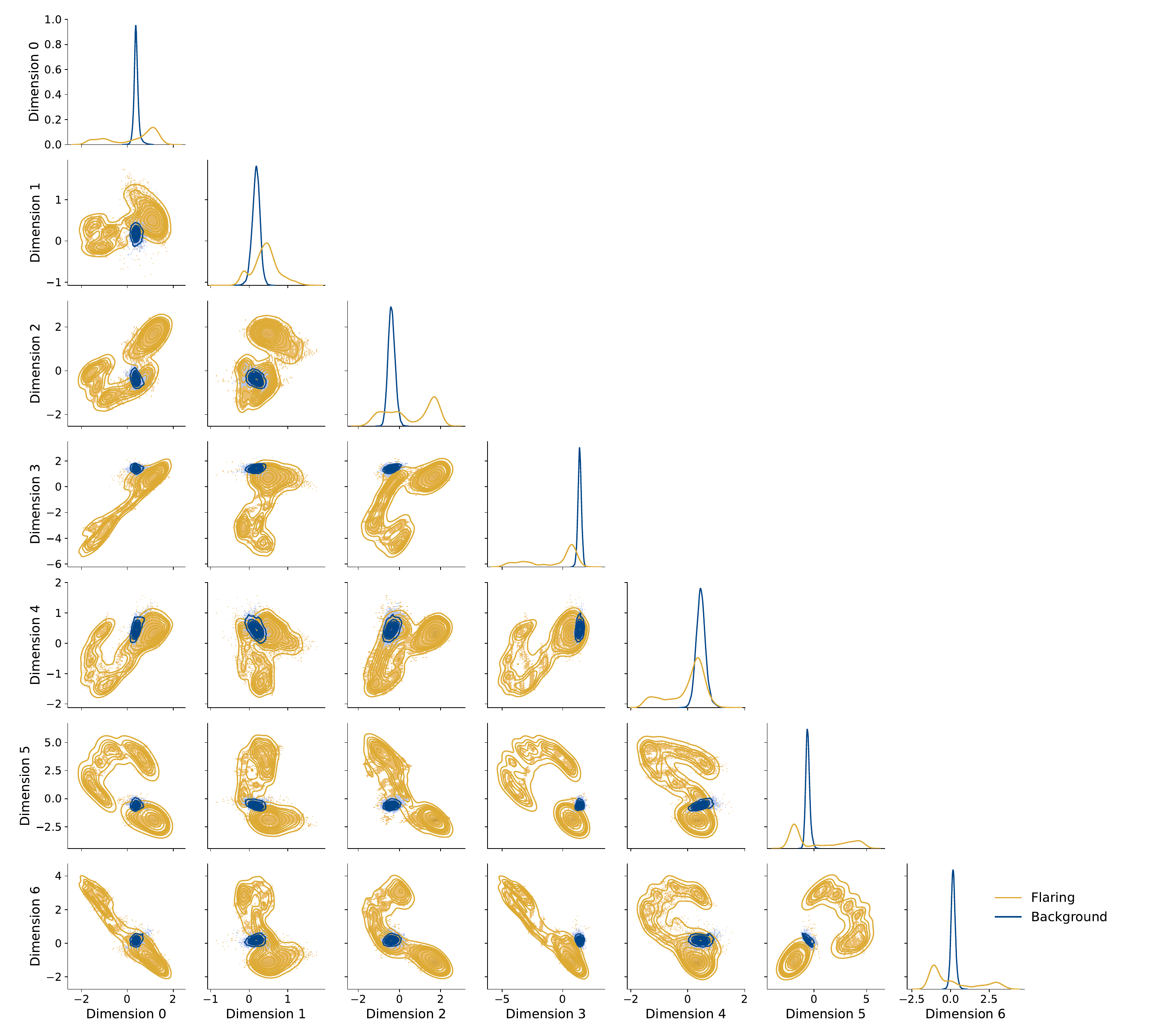}%
    \makebox[0cm][r]{%
    \raisebox{11cm}{%
      \includegraphics[width=.37\linewidth]{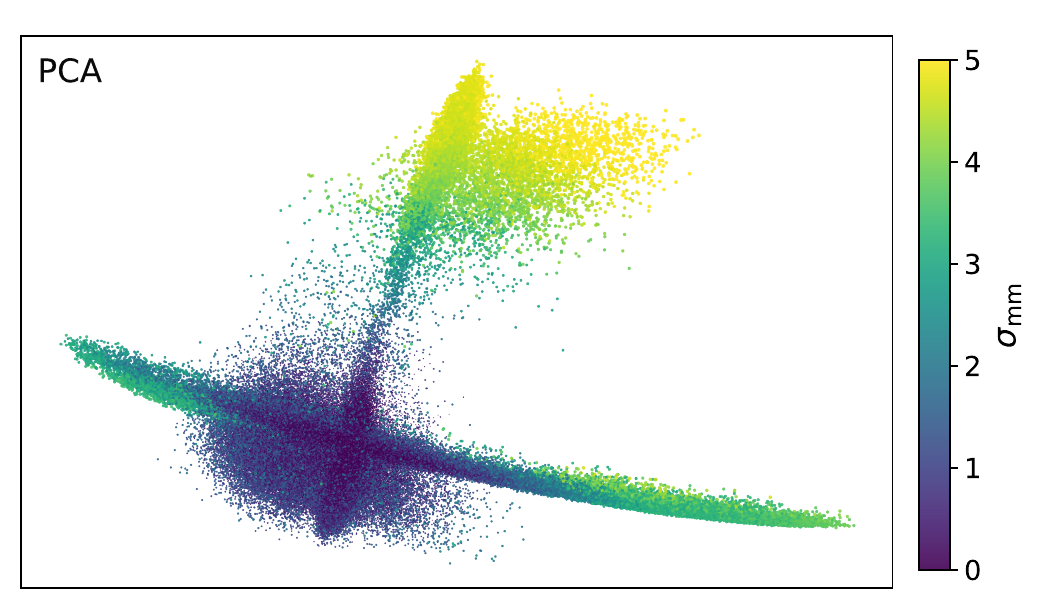}%
    }\hspace*{4.5cm}%
    }%
    \makebox[0cm][r]{%
    \raisebox{7cm}{%
      \includegraphics[width=.37\linewidth]{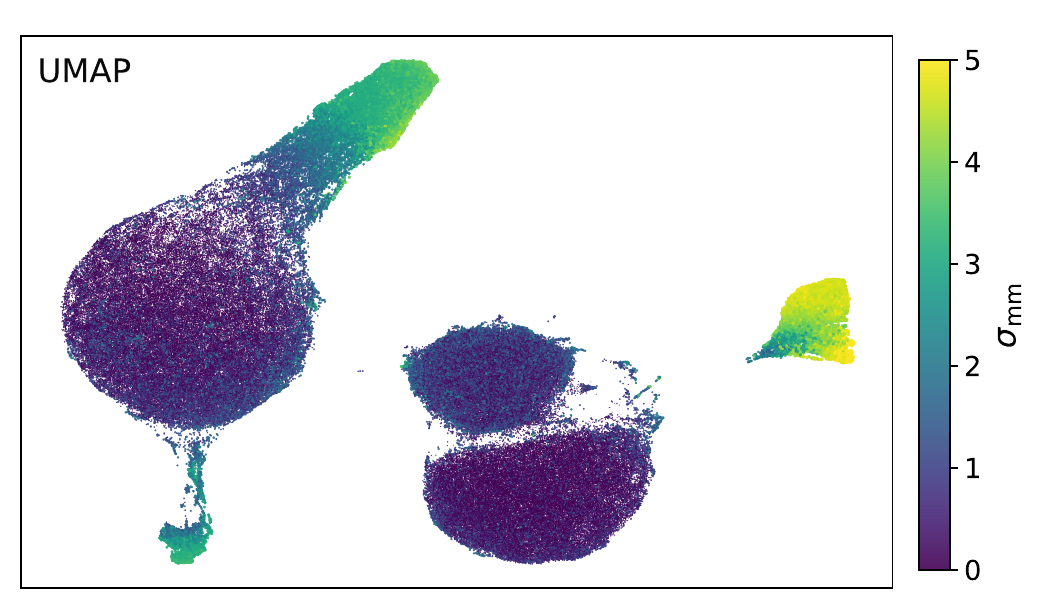}%
    }\hspace*{0.15cm}%
  }%
    \caption{Left: Bidimensional projections of the embedding space, showing the distributions of background (blue) and flaring states (yellow). 2D projections with PCA on the top left and UMAP on the top right. The color in the PCA and UMAP panels  indicates significance (based on the nominal embedding space). For visual clarity, marker size slightly increases with significance.}
    \label{fig:cluster-space-simulations}
\end{figure*}

\begin{figure*}[t]
    \centering
    \includegraphics[clip,width =
    .47\textwidth]{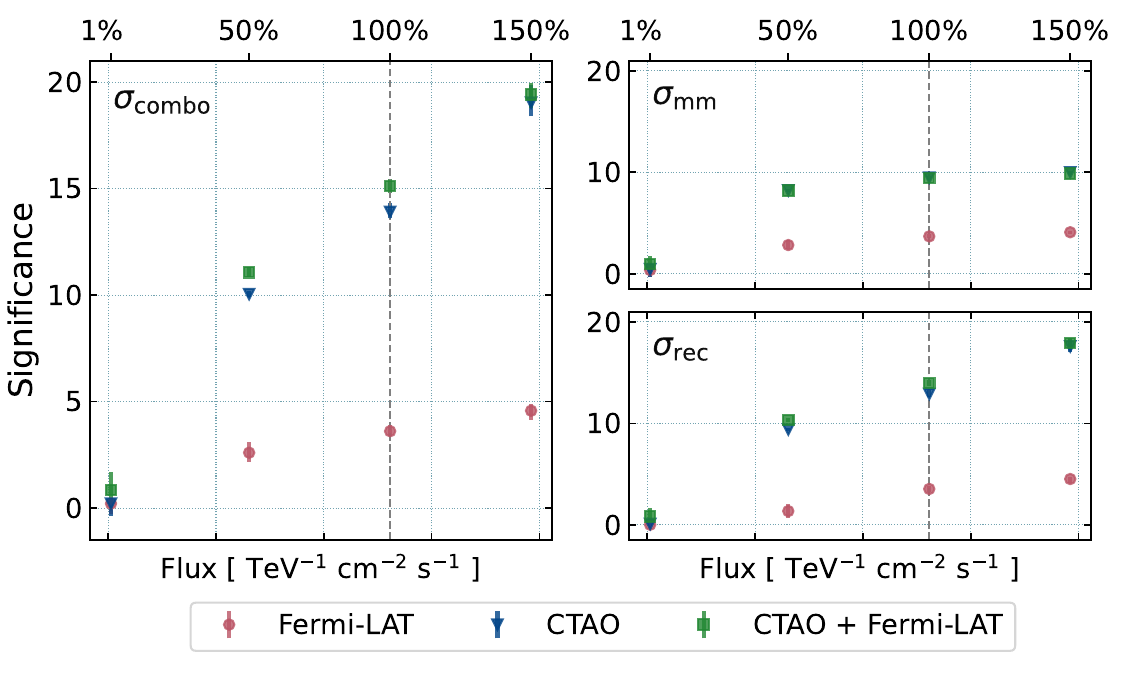}
    \quad \quad \quad 
	\includegraphics[clip,width =
	.47\textwidth]{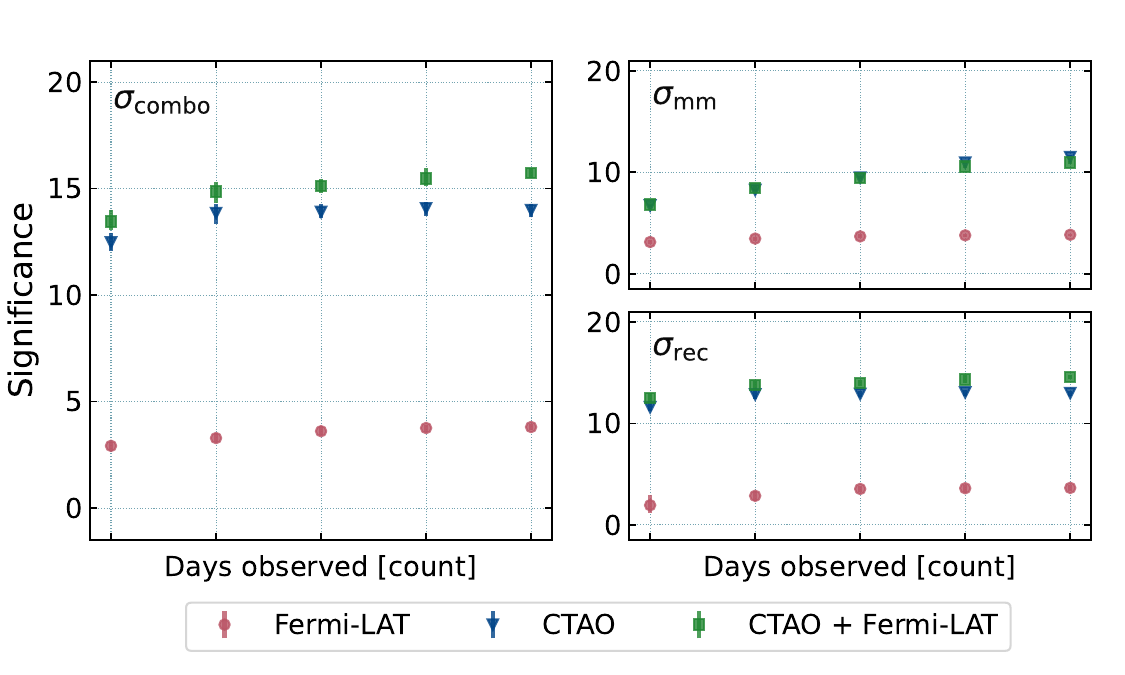}
    \caption{Left: Significance of flares, captured in three subsequent days, as a function of the simulated flux of the reference source,
    1ES$~$1215$+$303.
    The dashed line corresponds to the originally reported flux, $F_\mathrm{ref}$, from the 2017 flare of the source
    \citep[][]{Valverde_2020}{}{}. Right: Significance assuming
    the baseline flux of the source, $F_\mathrm{ref}$, as a function of the number of consecutive observation days. Different combinations of channels, \fermi and \cta, are considered,
    as indicated.
}
    \label{fig:simulation-results}
\end{figure*}

\begin{figure*}[t]
	\centering
	\includegraphics[trim={0cm 0cm 0cm 0cm}, clip, width =
	0.342\textwidth]{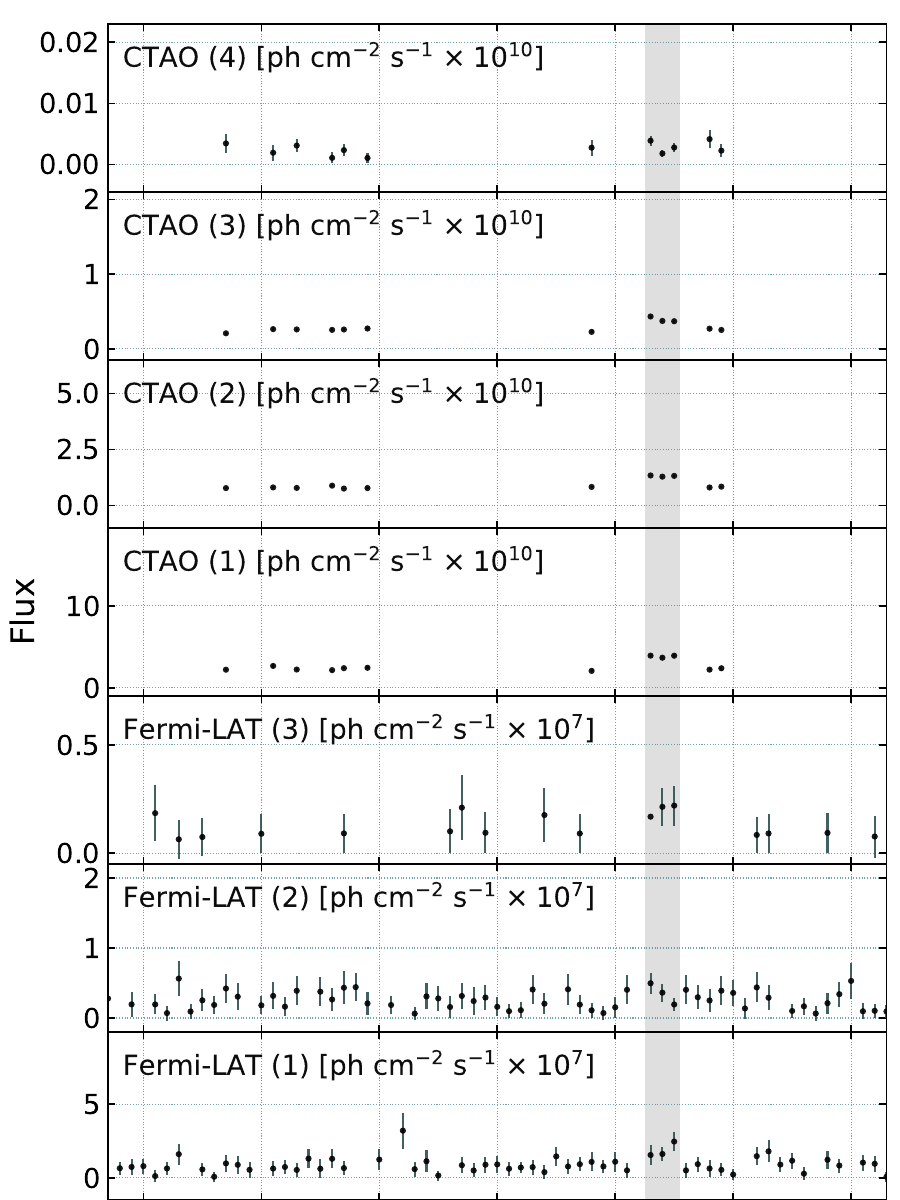}
	\includegraphics[trim={1cm 0cm 0cm 0cm}, clip, width =
	.32\textwidth]{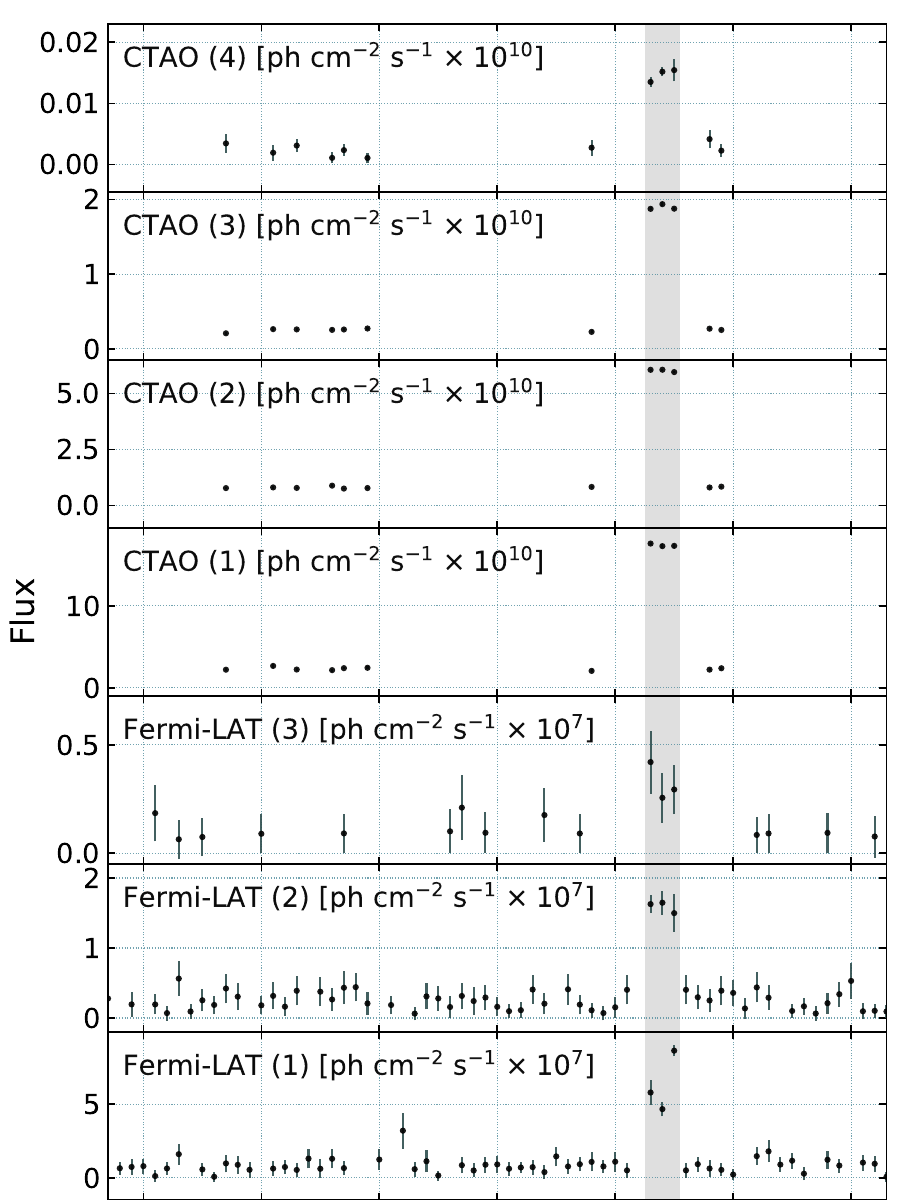}
	\includegraphics[trim={1cm 0cm 0cm 0cm}, clip, width =
	.32\textwidth]{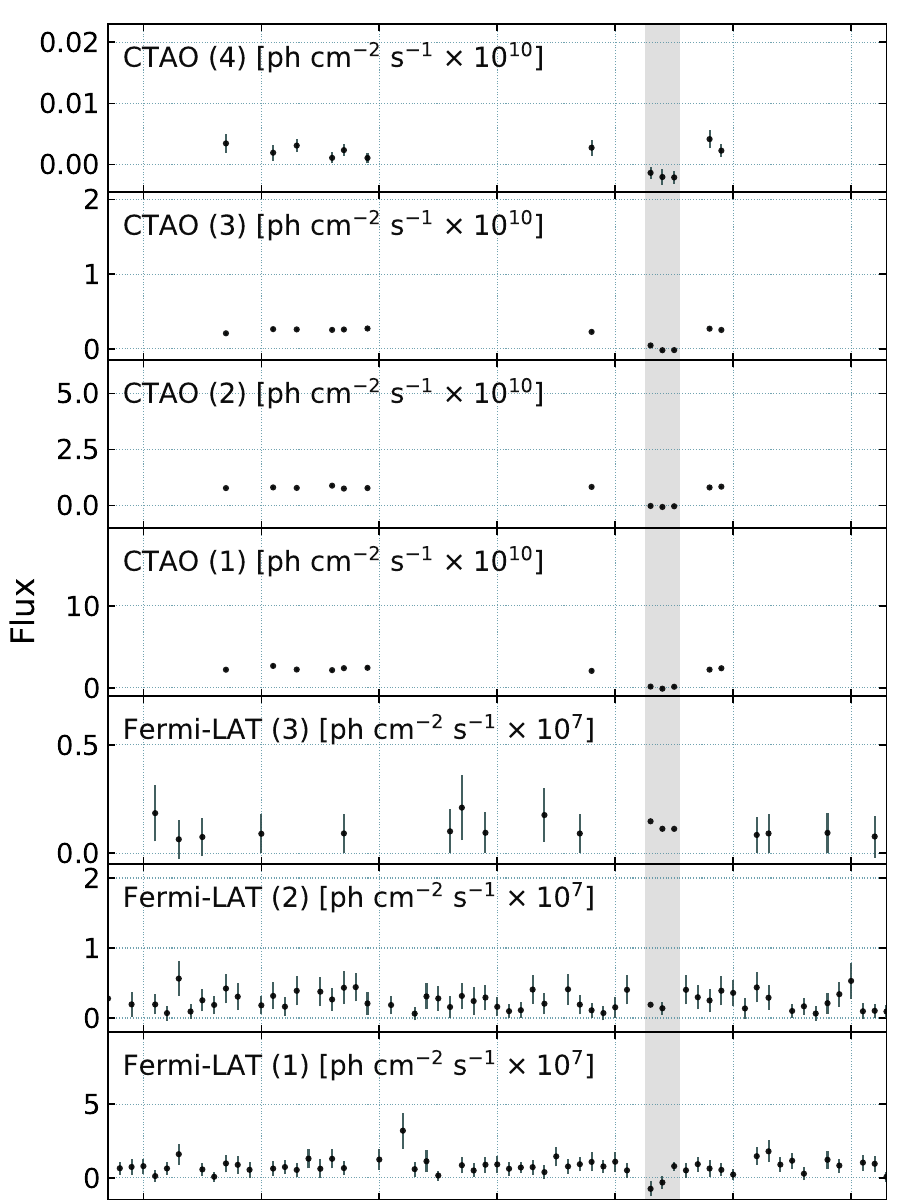} 

	\includegraphics[trim={0cm 0.cm 0cm 0cm}, clip, width =
	0.342\textwidth]{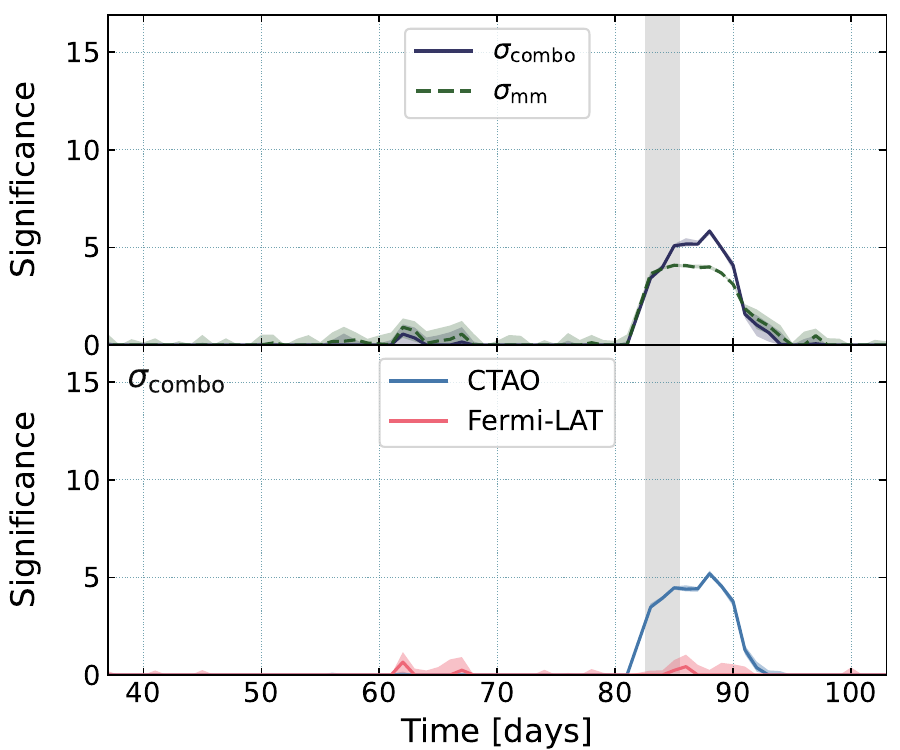}
	\includegraphics[trim={1cm 0.cm 0cm 0cm}, clip, width =
	.32\textwidth]{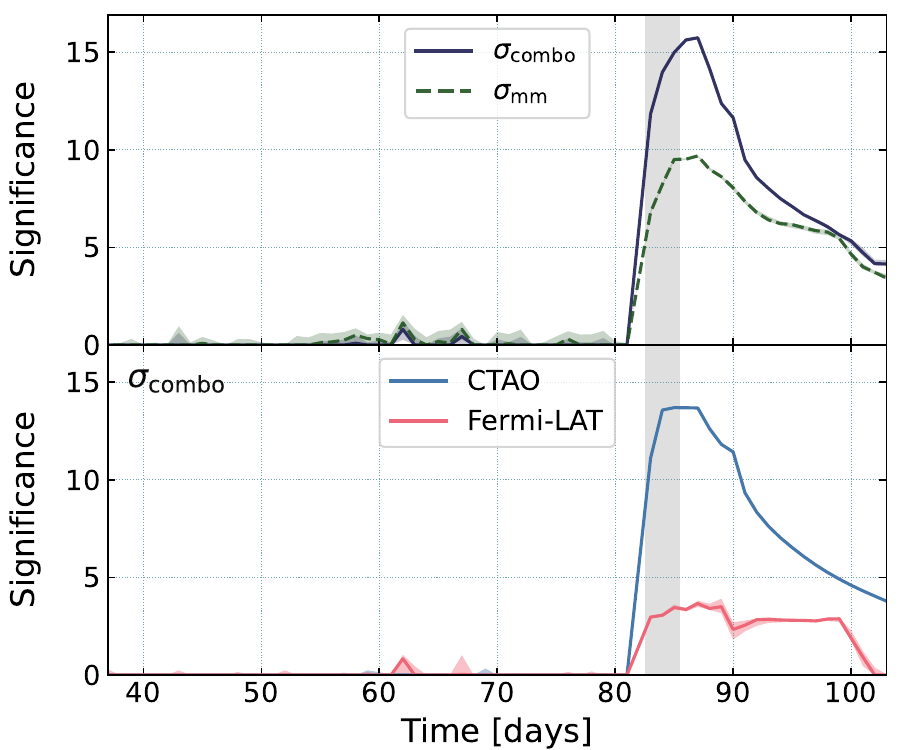}
	\includegraphics[trim={1cm 0.cm 0cm 0cm}, clip, width =
	.32\textwidth]{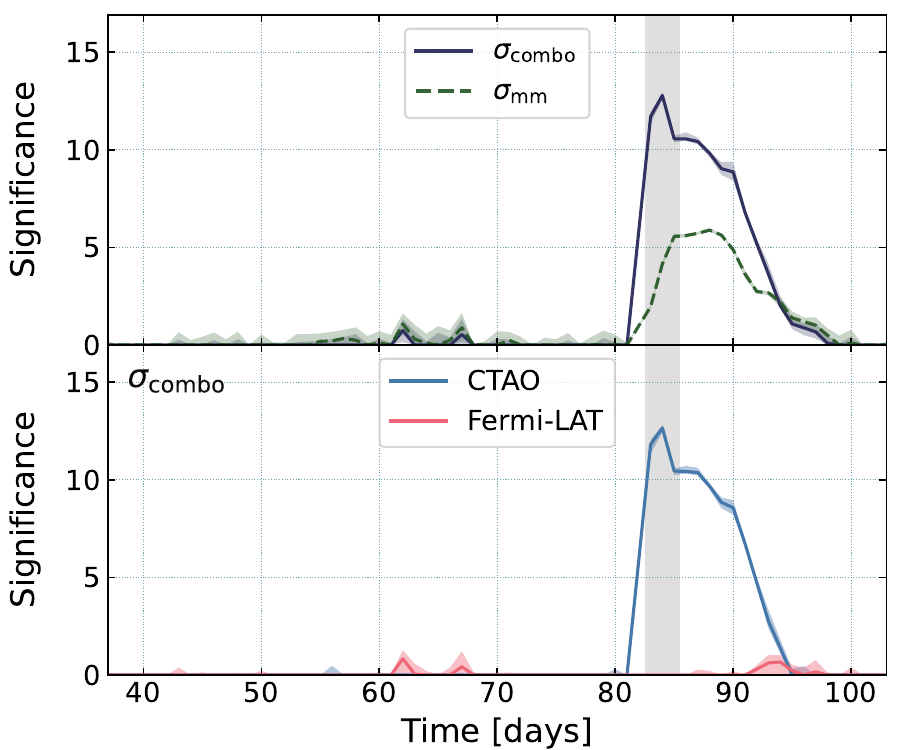}

	\caption{Illustration of injected flare templates, given scaling factors, 10\% (left), 100\% (middle) and  $-15$\% (right), applied to the baseline flux, $F_\mathrm{ref}$. The top panels show light curves in the various channels, as indicated. Energy bins for \cta (panel rows 1--4) and for \fermi (panel rows 5--7) are specified in \autoref{simdataset}.
    The second to last panel shows the 
    clustering significance, $\sigma_\mathrm{mm}$, and the 
    combined significance, $\sigma_\mathrm{combo}$. The lowest panel illustrates the
    contributions of individual channels to
    the combined significance. Flaring periods are highlighted as gray shaded regions.}
	\label{fig:simulation-lc_panels}
\end{figure*}

To study the performance of our pipeline, we first show that flaring and background states are distinguishable by their respective embedding vectors obtained from the autoencoder.
We then assess that the detection significance scales as expected with varying flare strengths, considering the expected time coverage within the \fermi and \cta channels. 

\subsubsection{Background and Flaring States as Embeddings}

We expect background and flaring states to be mapped into distinctive embedding clusters. To confirm this, we visualize the embedding vectors of our simulated dataset in 
\autoref{fig:cluster-space-simulations}.
In the bottom left, we show the distribution and pair-wise projections for each embedding vector dimension. Background states are shown in blue, and flaring states in orange. 
For illustration, we also show 2D projections of the embedding vectors using standard dimensionality reduction techniques to visualize complex data, namely Principal Component Analysis
\citep[PCA;][]{doi:10.1080/14786440109462720}{}{}
on the top, and Uniform Manifold Approximation and Projection
\citep[UMAP;][]{mcinnes2018umap}{}{} 
on the bottom.
Here, the color of each point shows the associated significance, with low significance (\ie data consistent with background states) shown in blue, and high-significance data shown in yellow.

It is apparent that background and flaring states are distinct, where some dimensions are more important for distinguishing between states than others. The linear (PCA) and nonlinear (UMAP) projections into 2D both show a notable difference between flaring and background states. A certain overlap between the two distributions is visible and expected, as some aspects of the data can be consistent between background and (weak) flaring states. 

Another observation is that flaring states cover a larger space compared to background states in most dimensions of the embedding.  The larger spread in the embedding vectors stems from the larger intrinsic variability of light curves in flaring states. The smaller variations in the background states are consequently mapped to a relatively  compact region. 
This is a desired outcome of the model, illustrating that the representations capture features of variability, which are the basis for anomaly detection.

\subsubsection{Significance and Flare Strength}
The detection significance is expected to scale with the strength of a flare. To study this, we show the significance as a function of flare strength in the three panels on the left of \autoref{fig:simulation-results}. The leftmost panel shows the combined significance, while the next two panels show the mixture model significance (top) and the significance of the reconstruction error (bottom).
In each panel, we show the respective significance for flares related to both \fermi and \cta, or flares exclusively affecting either \fermi or \cta.
The error bars illustrate the variance of the significance for the ten randomized realizations of each flare template.

We use scalings of the  strength of flares, ranging from 1\% to 150\% of $F_\mathrm{ref}$, given a fixed flare duration of three days.  
The dashed vertical line marks a scaling of 100\%. 
We expect very faint flares to be compatible with background states. This is the case, as can be seen by the low significance from flares with a flux of 1\%~$F_\mathrm{ref}$. 
With increasing strength, the significance rises.

The individual significance values from the mixture model and the reconstruction error are similar; both contribute to the combined significance. For strong flares, \ie scalings above 100\%, the significance from the mixture model appears to saturate. However, the significance of the reconstruction error continues to increase with the strength of flares. Consequently, the increase is also continuous in the combined significance. 

The highest significance is reached when a flare is present in both channels. When a flare is present only in \cta, the corresponding significance is slightly lower. Flares that only occur in \fermi result in significance values lower by a factor of a few. The only exception are faint flares (1\%~$F_\mathrm{ref}$), which always result in significance values close to zero. 

It is expected that the significance of \fermi-only flares is lower than \cta-only flares. This is because \fermi has larger flux uncertainties than \cta, and therefore the same flare appears less distinct from background variability in \fermi than in \cta. This is apparent by the fact that we obtain a significance ${>5\sigma}$ from a \cta-only flare with 10\% strength, whereas a \fermi-only flare with 50\% only shows a significance of around ${2.5\sigma}$.     
The significance of flares occurring in both channels is generally higher than the significance of a flare in only a single channel.

Besides flaring activity corresponding to elevated short-term states in observables, \framework is also sensitive to short-term decreases in input values. We show this in the right panel of \autoref{fig:simulation-lc_panels}. Here, a  flare is modeled by a  15\%~\fref decrease from background levels, which is clearly detected with significance $>12\sigma$ for \cta. However, this deviation is comparable with background fluctuations
for \fermi, given
the large intrinsic uncertainties in this channel. As expected, the significance increases when 
considering
additional channels that exhibit flaring activity. 

\color{black}
\subsubsection{Time Coverage}
Time coverage describes the number of time steps per channel that are affected by a flare. Coverage can greatly impact the probability for detection.
For weak flares, wider time coverage is required in order to identify anomalies. Conversely, 
even a single time step might suffice to confidently claim the detection of a strong flare.

We show the impact of time coverage on detection significance in the three panels on the right of \autoref{fig:simulation-results}. Here, we simulate observations of a flare with 100\%~$F_\mathrm{ref}$ and time coverage between one and five time steps. We show the combined significance (third column) and the contributions of the individual significance components (last column), also differentiating between detections in only some, or in all channels. 

Most notably, the same trends are observed  related to 
independent detections in \cta or 
in \fermi,
as described in the previous section.
Flares detected only in \fermi generally result in lower significance. This is again due to the \fermi light curve being intrinsically noisier than the \cta data. Furthermore, we find that a second detection channel, even if much less significant on its own, can increase the overall significance.

Flares detected in \cta or in \cta and \fermi reach a high combined significance  with a single high-flux observation. With each additional day of coverage, the significance increases more slowly and begins plateauing beyond three days. 
Flares only present in \fermi channels on the other hand yield a low combined significance at a coverage of one day; the significance then increases continuously, reaching  a plateau beyond four days. This is consistent with the
expectation that a
weak flare would require longer coverage in order to become significant. 

One may observe that the reconstruction significance (bottom right panel) dominates the overall significance for all cases where a flare is detected in \cta. This can be explained by the makeup of the training dataset;
it consists of random, short-term fluctuations with different amplitudes and durations. Due to the intrinsic sparsity in the \cta channels, only a few cases of fluctuations having coverage of more than one time steps appear in the training data. Limited training data for this particular pattern of multi-day flares results in potentially greater reconstruction errors.
Conversely, the \fermi training data, due to its much denser sampling, contains relatively more multi-day fluctuations. The corresponding reconstruction significance is therefore much less dominant.
The impact of incomplete training data is mitigated by the complementary use of both 
$\sigma_\mathrm{mm}$
and
$\sigma_\mathrm{rec}$.

\subsubsection{Time Profile of a Detection} We expect the significance to rise in the presence of new flaring data, and to then gradually fall as physical time passes. 
We study this behavior in \autoref{fig:simulation-lc_panels}. We show three zoomed in timelines with light curves and their corresponding significance values for a weak upward flare (+10\%~$F_\mathrm{ref}$) on the left, and a strong upward flare (+100\%~$F_\mathrm{ref}$)  in the center. On the right, we show a moderate downward flare ($-15\%$~$F_\mathrm{ref}$).
The lowest panel illustrates the contribution to the detection significance when considering individual channels. 
The flare is highlighted as the gray shaded region. 

As expected, the significance starts rising as soon as the first flaring time step comes into play. It reaches a plateau just after the end of the flaring period, and then decays with time. 
In the case of the weaker flare on the left side of  \autoref{fig:simulation-lc_panels}, there is a small peak after the flaring period. Here a time step in the topmost channel exhibits comparable strength to the preceding flare. The continuously elevated state on this timescale is incompatible with background, resulting in high significance. Thereafter, the significance falls, as the variance of the following data returns to background levels.
The example of injecting a negative fluctuation illustrates that \framework is agnostic to the topology of potential flares.

\section{Historical Blazar study}
\label{sec:realdata}

After evaluating our framework on simulations, we proceed to analyze historical data and flares of the blazar BL Lacertae. We assess the effectiveness of our framework by comparing with the reactions of the real-time \mwl community. In this context, we make use of alert systems, such as the General Coordinates Network (GCN) and the Astronomer's Telegram (ATel). ATels can be published with a delay of up to a few days. We therefore compare the date \framework detects an anomaly
to the time of observation, rather
than to the date of publication.

\subsection{Dataset}\label{sec:bllacdataset}
We obtained \mwl data of the blazar BL Lacertae from a number of instruments, operating in different parts of the electromagnetic spectrum.
The data span a time range of $\sim$4,000 days, MJD 55,170 -- 59,200 (between December 2009 and December 2020). Observation times are rounded to a timescale, ${\gammatimeaggr = 1}$~days.
A combined \mwl light curve for these data is shown in the top panels of \autoref{fig:blac-lc-and-results}.
A detailed description of the individual datasets and of the related processing and analysis steps is given in \autoref{App:MWL_DATA}.

This light curve includes data from the following observatories.
In the very high energy (VHE) band between $100~$GeV and $30~$TeV, we use data from VERITAS, an array of four imaging atmospheric Cherenkov telescopes in Arizona \citep[][]{Holder2008}{}{}. While VERITAS has published several distinct emission episodes of BL Lac \citep[][]{arlen_2013_bllac, feng_2017_bllac, veritas_2018_bllac}, the complete dataset remains private. The collaboration has kindly allowed us to explore all available data for our analysis. Accordingly, we use the flux of the source as input to our pipeline. For visual purposes only (\autorefs{fig:blac-lc-and-results}, \ref{fig:result-highlights} in the following), we mask the data. Instead of flux, we
show the corresponding excess photon count rate, slightly modified by random noise. Though the data presented in the figures are in arbitrary units, they track the relevant changes to the measured flux of the source. They can thus
be used to appreciate the performance of our method, as discussed below.

For energies between $100~$MeV and $300~$GeV we use data taken by the Large Area Telescope (LAT) onboard the \fermi spacecraft \citep[][]{atwood_2009_fermi_lat}{}{}, divided into the same three energy bins as before
(100 -- 669~MeV,
669~MeV -- 4.48~GeV, and
4.48 -- 300~GeV).
In the X-ray band, we include observations of BL Lac taken by the X-Ray Telescope (XRT) onboard the \textit{Swift} satellite \citep[][]{burrows_2005}{}{}, divided into soft ($0.3-1.5~$keV) and hard bands ($1.5 - 10~$keV).

Optical r-band data include observations taken with the Tuorla optical telescope in Finland \citep[][]{nilsson_2018}{}{}; observations taken with the Palomar Transient Factory (PTF) monitoring program \citep[][]{law_2009}{}{} in San Diego County, California; and with the successor instrument, the Zwicky Transient Facility \citep[][]{bellm_2019a, bellm_2019b}{}{}.
The latter two observatories also provide g-band magnitudes of the source.

Finally, we use radio/microwave data at wavelengths of 0.87$~$mm and 1.3$~$mm, taken by the Submillimeter Array (SMA) located on Mauna Kea in Hawaii \citep[][]{gurwell_2007}{}{}.

Data from VERITAS, \fermil, and SMA are in units of flux; data from Tuorla and ZTF/PTF are in magnitude (Mag.), and data from \textit{Swift}-XRT consist of count rates.
We decided against performing an advanced X-ray analysis, which would include using a spectral model and fitting the neutral hydrogen column density. Instead, we use the lower-level data product, provided through the \textit{Swift} online analysis tools (see also \autoref{App:MWL_DATA}).
By design, \framework is agnostic to using different types of input across channels.
We approximate the uncertainties on inputs by sampling; we use Gaussian distributions, having standard deviations corresponding to the quoted uncertainties.

In order to create a training sample that exclusively comprises  quiescent states, we first exclude data exhibiting exceptional source activity.
The definition of exceptional activity as part of data cleaning determines
the sensitivity threshold of the model.
We decide to first manually exclude clearly identifiable high states from the light curves.
In particular, we remove flaring states based on  VERITAS data surrounding MJDs 55,725, 57,200, 57,675, 58,100 and 58,600. We exclude any data taken within 50 days of the respective VHE flares from all channels.

We proceed to perform algorithmic cleanup. We exclude all data  for a particular time period, where at least one of the channels exceeds a given cutoff.
Cutoff values are defined in terms of the variance of a channel.
Specifically, we exclude data above a maximum flux threshold in the VERITAS, \fermil, and SMA channels; above a maximum count rate in the \textit{Swift} channels; and below a minimum magnitude
for Tuorla and ZTF/PTF.

\subsection{Configuration Parameters and Training}

The overall model and parameterization of training are analogous to those used for the simulation study.
However, we increase the  complexity of the model, as well as the regularization parameter.
While we do not perform an exhaustive hyperparameter optimization, we manually adjust some key parameters, in order to minimize underfitting and overfitting.
Specifically, we double the size of the
RNN
layers, double the dropout rate to $20\%$, and increase the L2 norm to $0.001$.

\subsection{Results}

\begin{figure*}[t]
    \centering
    \includegraphics[trim={0cm 0cm 0cm 0cm}, clip, width =
    0.97\textwidth]{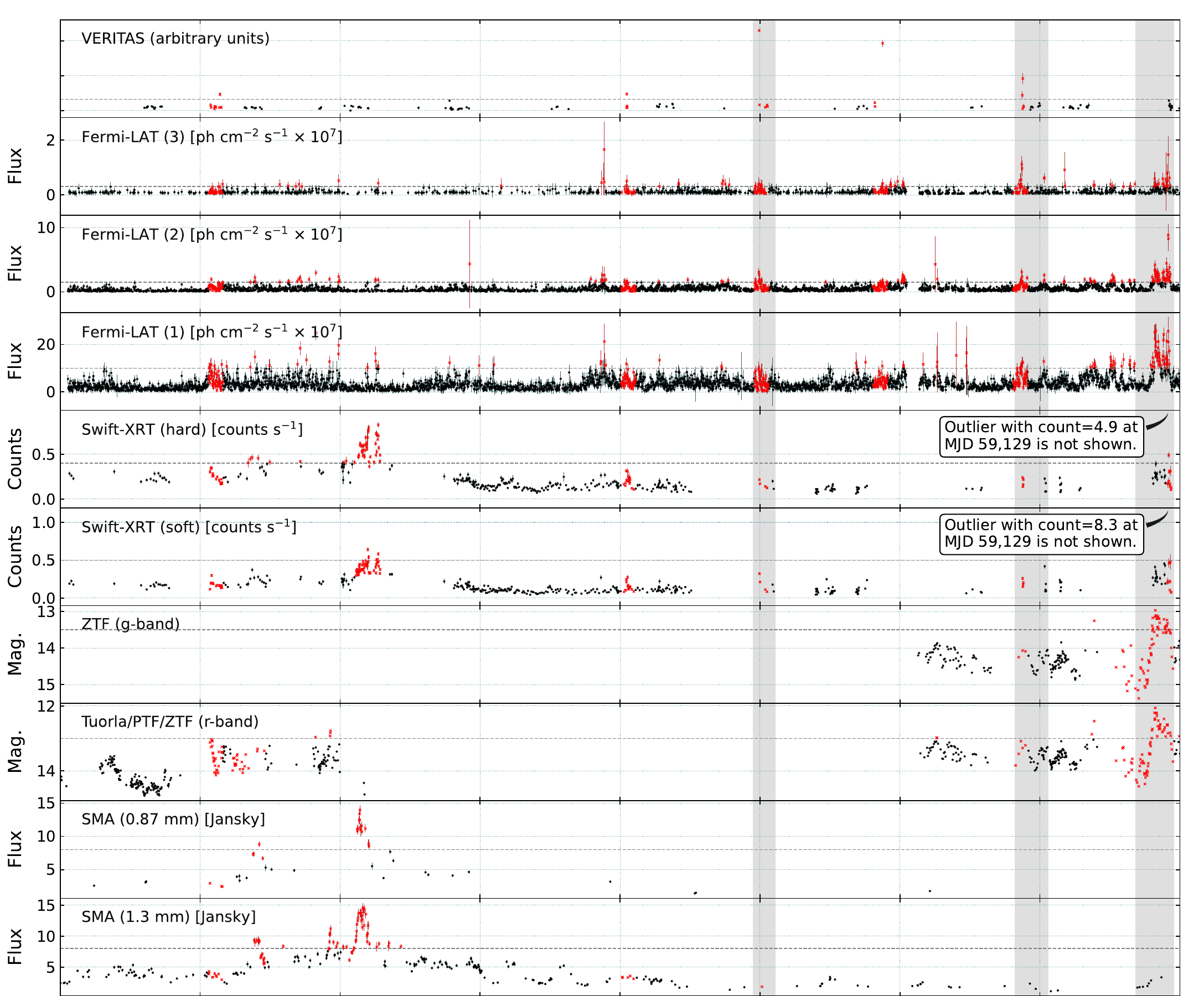}
    \includegraphics[trim={0cm 0.5cm 0cm 0cm}, clip, width =
    0.97\textwidth]{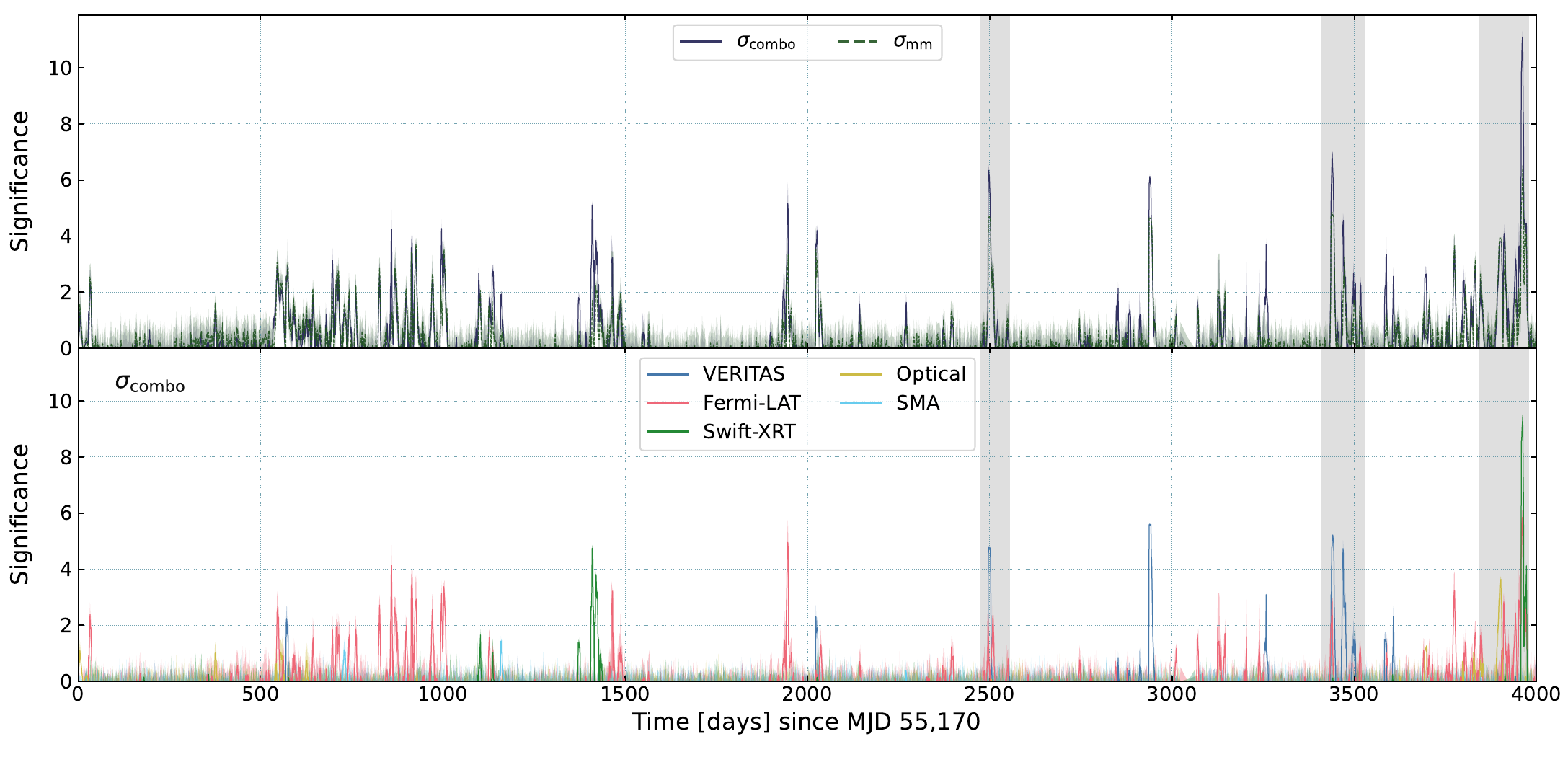}
    \caption{Overview of the \mwl BL Lac dataset and the corresponding significance of flares, $\sigma_\mathrm{mm}$ and $\sigma_\mathrm{combo}$. The lowest panel illustrates the contributions of individual channels to the combined significance.
    Data marked with red crosses are excluded from
    the training and calibration samples.
    Horizontal dashed lines indicate the associated background cuts.
    Gray shaded regions mark the time ranges highlighted in \autoref{fig:result-highlights}. Both \textit{Swift}-XRT channels contain an outlier,  hidden for readability.
    }
    \label{fig:blac-lc-and-results}
\end{figure*}

\begin{figure*}[t]
    \centering
    \includegraphics[trim={.0cm 0cm 0cm 0cm}, clip, width =
    0.342\textwidth]{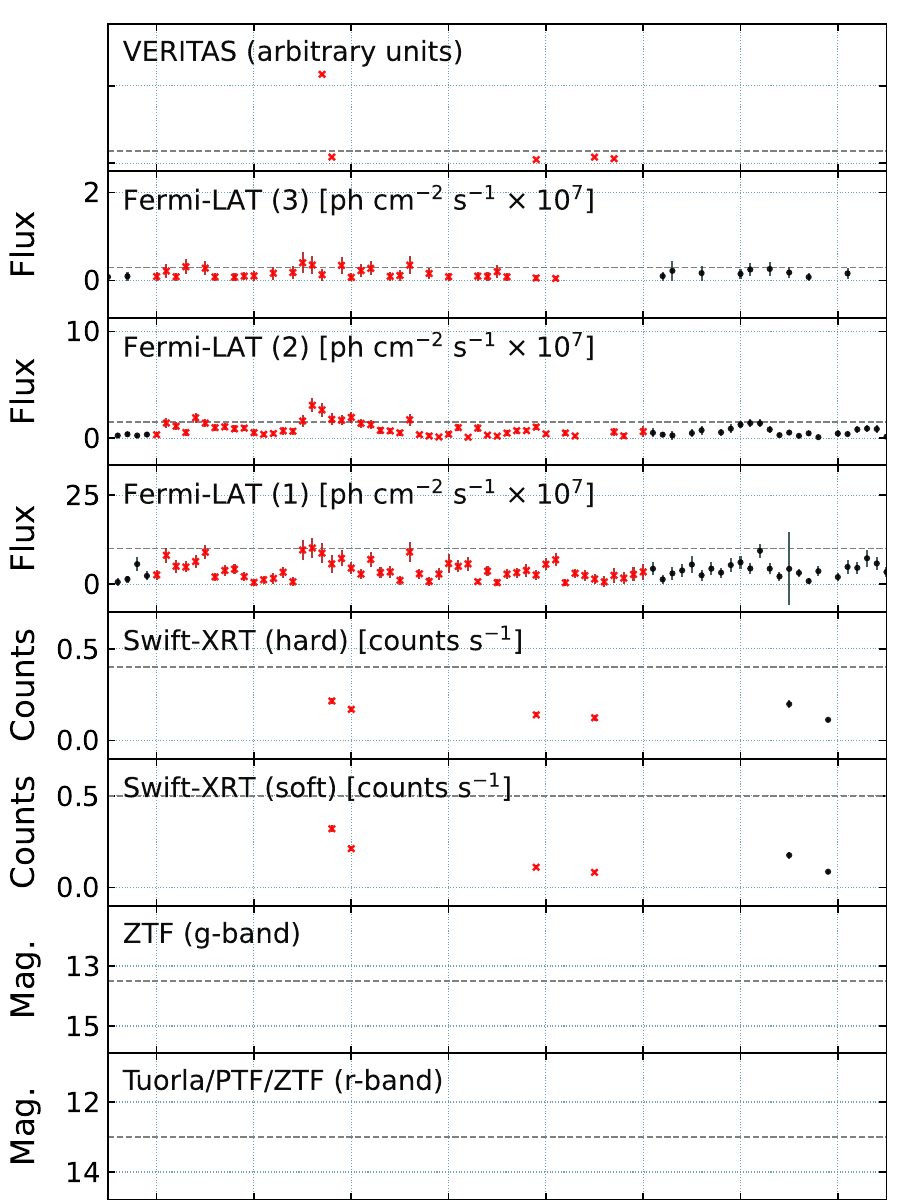}
    \includegraphics[trim={1cm 0cm 0cm 0cm}, clip, width =
    0.32\textwidth]{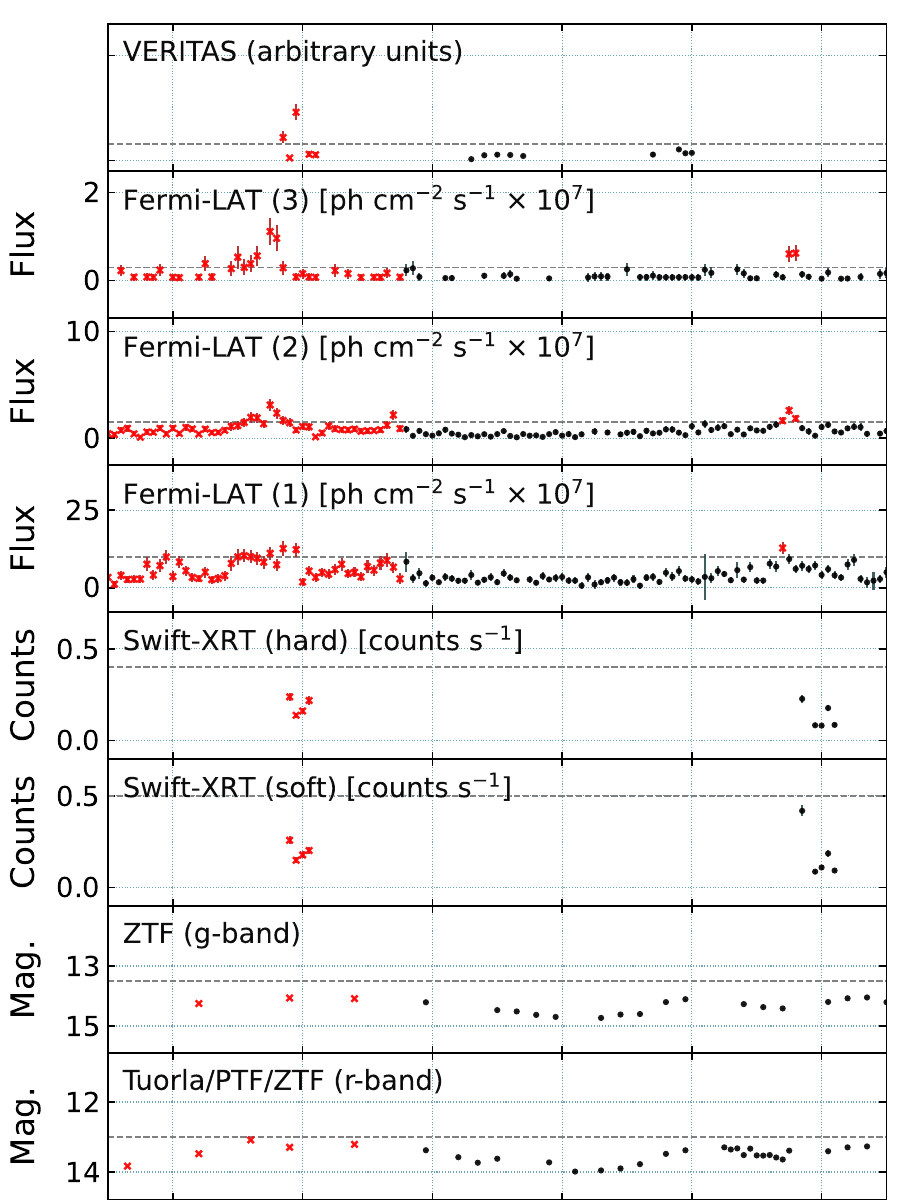}
    \includegraphics[trim={1cm 0cm 0cm 0cm}, clip, width =
    0.32\textwidth]{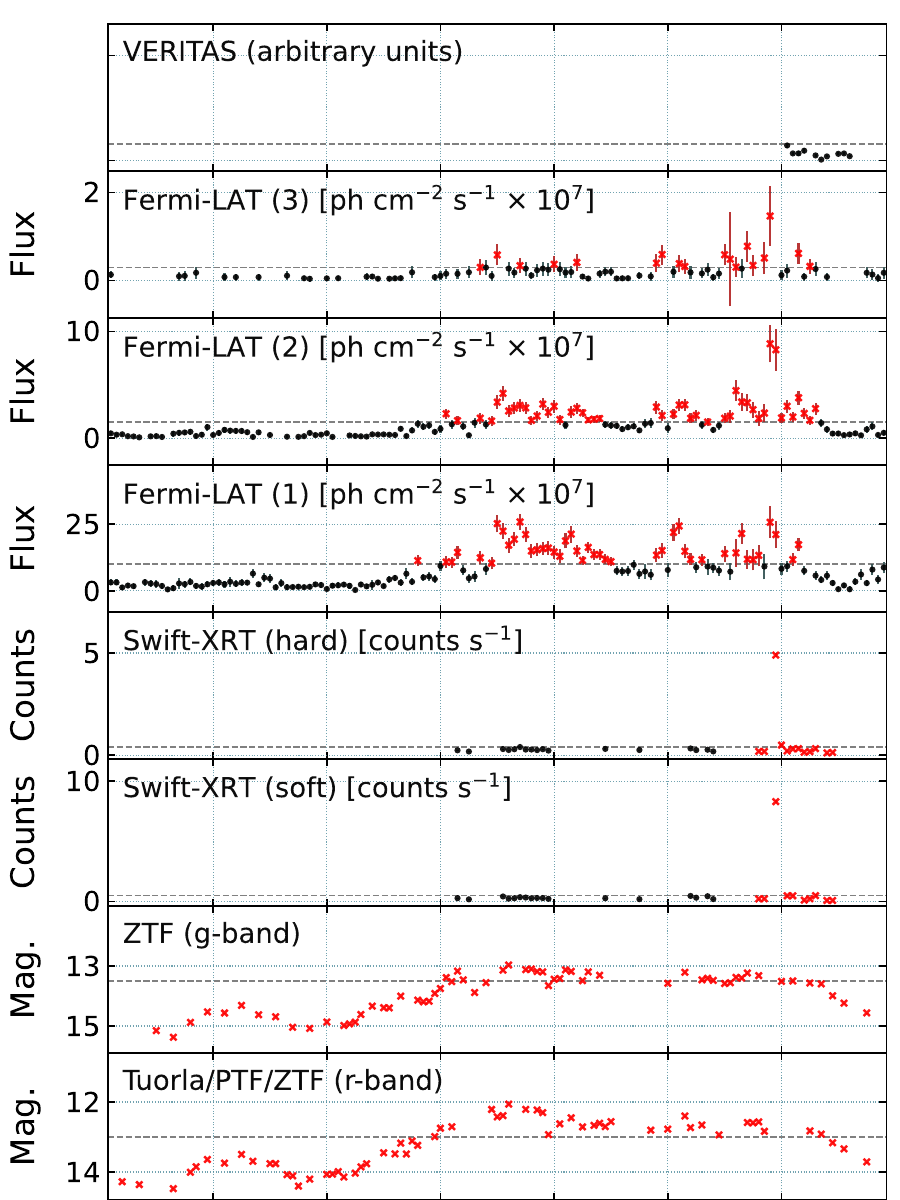}

    \includegraphics[trim={.0cm 0cm 0cm 0cm}, clip, width =
    0.342\textwidth]{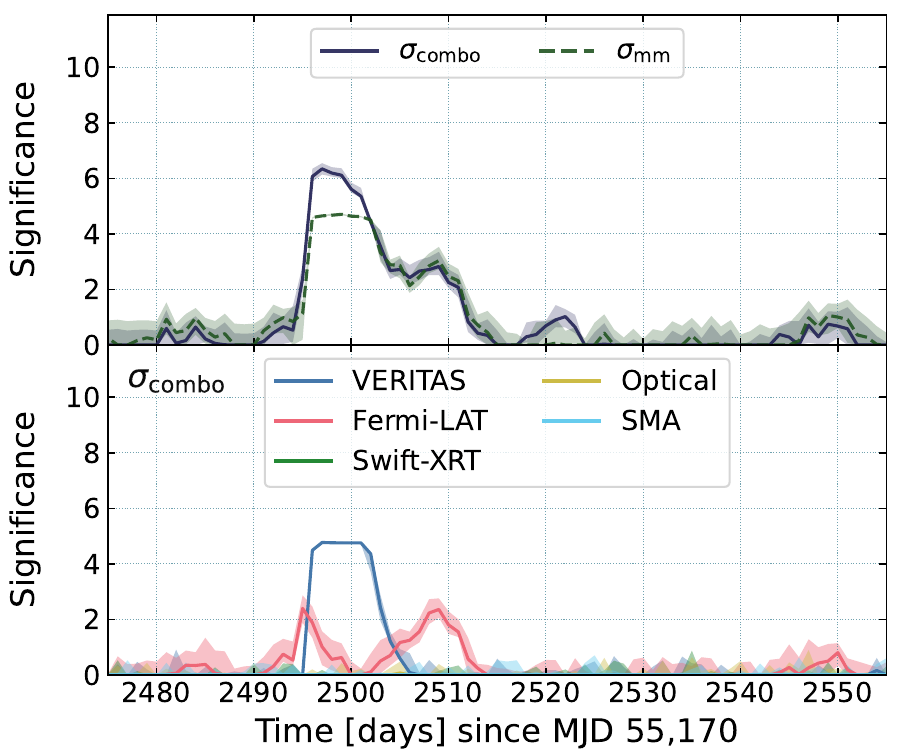}
    \includegraphics[trim={1cm 0cm 0cm 0cm}, clip, width =
    0.32\textwidth]{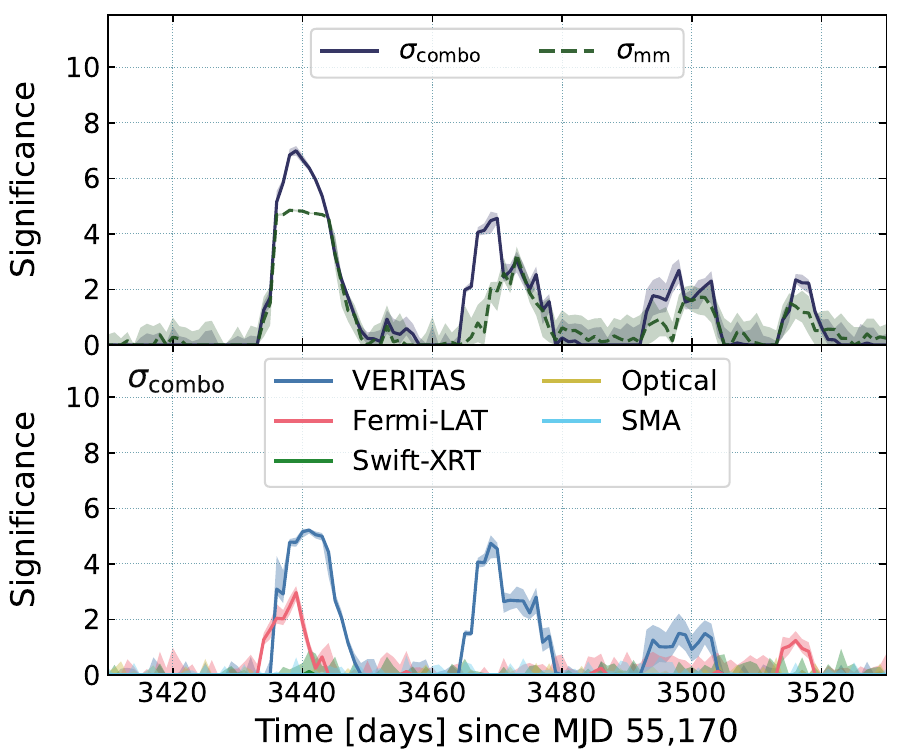}
    \includegraphics[trim={1cm 0cm 0cm 0cm}, clip, width =
    0.32\textwidth]{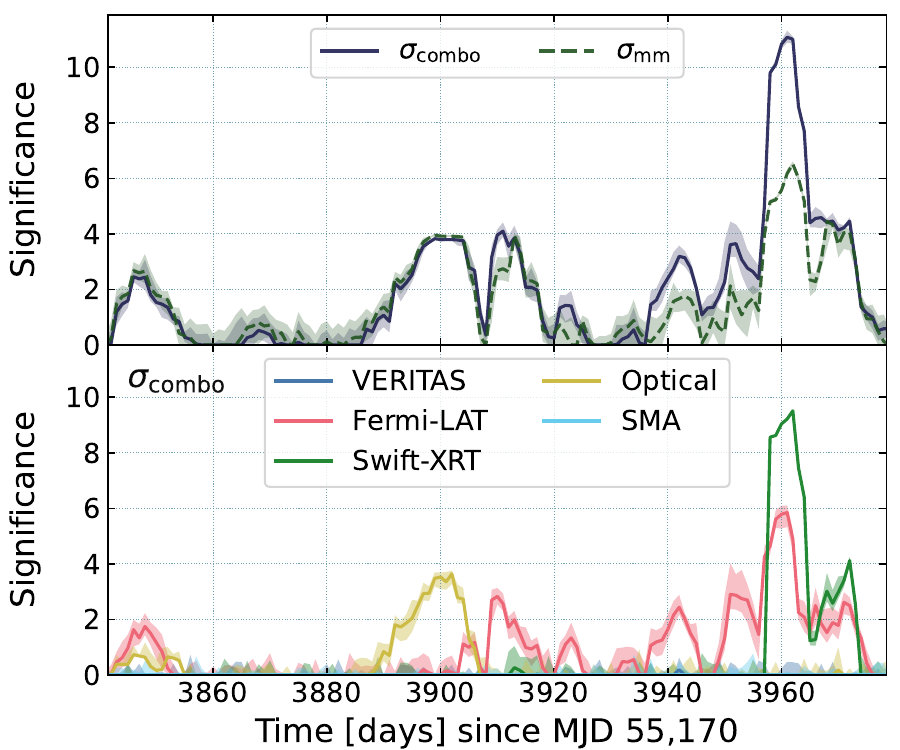}
    \caption{Highlighted intervals related to high-significance detections in the BL Lac dataset. (See caption of \autoref{fig:blac-lc-and-results}.)}
    \label{fig:result-highlights}
\end{figure*}

\begin{figure}[t]
    \centering
    \includegraphics[trim={0cm 0cm 0cm 0cm}, clip, width
    =\columnwidth]{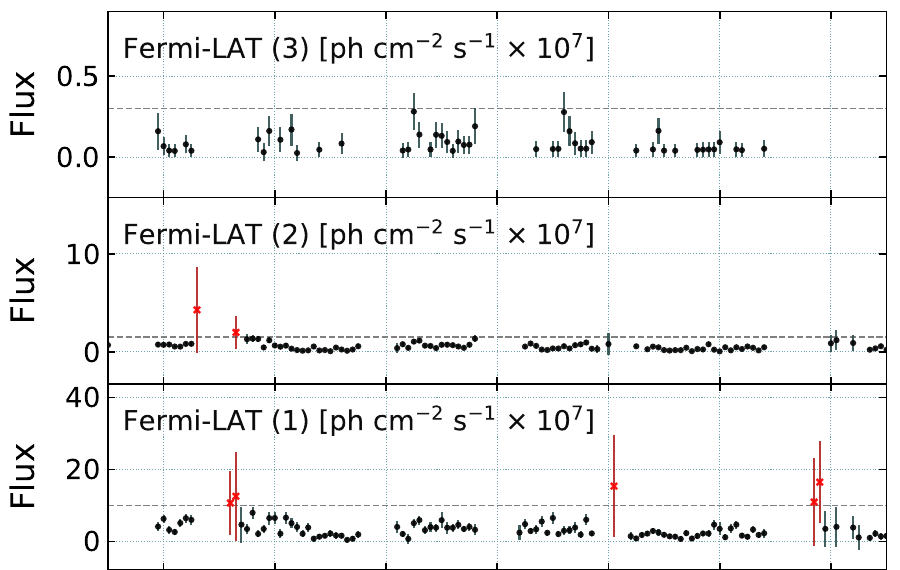}

    \includegraphics[trim={0cm 0cm 0cm 0cm}, clip, width
    =\columnwidth]{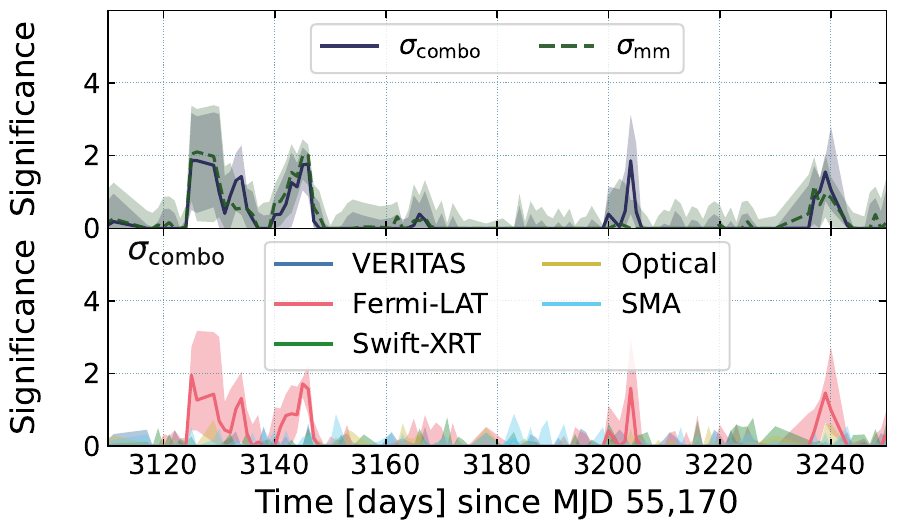}
    \caption{Highlighted interval from the BL Lac dataset. (See caption of \autoref{fig:blac-lc-and-results}.)
    The high significance can be traced back to the \textit{Fermi-LAT} channel.
    Large uncertainties on these data yield correspondingly high uncertainties on the detection significance. This can be interpreted as a likely false positive signal.
    }
    \label{fig:result-uncertainty}
\end{figure}

In \autoref{fig:blac-lc-and-results}, we show the full \mwl light curve of BL Lac used in this work.
The outcome of the data cleaning process is indicated, where all excluded  time periods are highlighted by red crosses.
For context, the automatically derived  cleaning cut-offs are shown as horizontal dashed lines.
Time intervals highlighted in gray are shown separately in \autoref{fig:result-highlights}.
(The SMA channels are hidden for readability, as they do not contribute notably to the significance for detection.)
The two panels on the bottom  show the combined significance, and the clustering significance for all channels. The individual contributions to the significance from each channel are illustrative, as discussed above.

In the following, we refer to $t$ as the time (number of days) since MJD 55,170.

\subsubsection{Performance on Realistic Light Curves}

The data used here are much more complex than the simulated light curves investigated in \autoref{sec:simulation}.
We utilize a total of ten channels. These cover a wide range of the electromagnetic spectrum, having differing observing cadences and occasional extended gaps.
The 4,000-day light curve contains multiple flaring periods across different channels.
From the corresponding significance, one can clearly distinguish episodes of high activity from quiescent periods.
This shows that \framework is  able to distinguish between different blazar states.

The first flare contained in this light curve is a VHE flare, detected by VERITAS at $t = 570$ \citep[][]{veritas_atel_bllac_2011}. This flare is fainter than subsequent VHE flares, and is detected by \framework at $\sim 2.5\sigma$.
Three of the highest-significance peaks ($t = 2,\hspace{-1pt}500$, 2,900, and 3,400~days) coincide with \veritas flares, each  reaching ${\sim6\sigma}$.  A close-up view of the first peak is shown in the left panel of \autoref{fig:result-highlights}.
The first and third of these flares are accompanied by a flare detected by \fermil, as indicated by the individual channel contributions to the significance.

In general, we find elevated significance during the parts of the light curve that were excluded for training.
The main exception to this is the significance peak at $t \sim 1,\hspace{-1pt}450$.
At first glance, the surrounding light curve does not show increased variability.
The signal here is mostly driven by the reconstruction significance. The lower panel reveals that the largest contribution comes from the \textit{Swift}-XRT channels.
This can be explained by the fast transition from a clear high state into a low state. The transition is first observed at $t\sim1,\hspace{-1pt}300$, and continues after a gap in \textit{Swift} coverage.

The SMA data yield low-significance detections with a $\sim 1\sigma$ spike at $t\sim740$ and a $\lesssim 2\sigma$ spike at $t\sim1,\hspace{-1pt}250$, both following fast transitions between activity states in the respective channels.
The two peaks preceding the $\sim 2\sigma$ detection at $t = 1,\hspace{-1pt}250$ do not result in pronounced signals.
This is presumably due to the timescale of our analysis being days. The respective evolution of the SMA light curves is slower; it also exhibits
relatively high intrinsic variance, which increases the corresponding sensitivity threshold for anomalous behavior.

\subsubsection{Uncertainty Propagation}
In \autoref{fig:result-uncertainty},
we illustrate the impact of input uncertainties on the output of our model.
We show a zoom in of the \fermil channels for $t \sim 3,\hspace{-1pt}200$. Here,
multiple data points have relatively high values, but these are
also characterized by large uncertainties. While the corresponding median significance increases to $\sim2\sigma$,
the related uncertainty is quite high.
We can thus infer that the elevated state is likely a false-positive signal.

\subsubsection{Real-time Detection and Prediction of Increased Multiwavelength Activity}

In this section, we compare the performance of \framework with the response of the \mwl community to two historical flaring episodes.
A summary is provided in \autoref{tab:mwlresponses}; we show the light curves of these flares in the two panels on the right of \autoref{fig:result-highlights}.

The panel at the center shows a VERITAS flare between $t = 3,\hspace{-1pt}436$ and~3,438, which is easily identified by \framework at $\sim5\sigma$. Additionally, the flux in the \fermi channels, most notably in the highest-energy channel, \fermil 2, starts increasing around $t = 3,\hspace{-1pt}430$.
This precursor is also picked up by \framework, as we can see in the bottom panels. One can observe an increase in significance, driven by \fermi. For the initial flux increase in the \fermi channels, \framework yields a combined significance of $\lesssim 2\sigma$ around $t = 3,\hspace{-1pt}435$. On $t = 3,\hspace{-1pt}434$, this trend was noted by the \fermil collaboration, as reported in an ATel on the following day \citep[][]{fermi_atel_bllac_2019}.
This ATel was then used to trigger observations with VERITAS and the MAGIC telescope at $t = 3,\hspace{-1pt}436$;  VHE flaring activity was found and subsequently reported by \cite{magic_atel_bllac_2019}.

After this flare, subsequent observations with VERITAS around $t = 3,\hspace{-1pt}470$ and 3,500 reveal a return to a low-activity state. For these observations, our framework yields a significance of $\sim4\sigma$ and $\sim2.7\sigma$, respectively, driven entirely by the VERITAS channel. This shows that abrupt transitions from high to low states can also produce a significant detection, provided the low state is observed shortly after the high state.

\begin{table*}
    \centering
    \footnotesize
    \begin{tabular}{p{1.6cm} @{\hspace{1.2em}} p{3cm} @{\hspace{1.2em}} p{3.2cm} @{\hspace{1.2em}} p{8.6cm}}
    \toprule

    $\mathbf{t}$ \textbf{[days]} & \textbf{Event} & \textbf{Framework response (significance at time)} & \textbf{Historical community response} \\ \midrule

    $570$ & VHE flare & $2.5\sigma$ at ${t = 570}$ & Detected at ${t = 570}$ and reported in ATel at ${t = 571}$ by VERITAS \citep[][]{veritas_atel_bllac_2011} \\

    2,496 & VHE flare & $6\sigma$ at ${t = 2,\hspace{-1pt}496}$ & Detected at ${t = 2,\hspace{-1pt}496}$ and reported in ATel at ${t = 2,\hspace{-1pt}496}$ by VERITAS \citep[]{veritas_atel_bllac_2016} \\

    3,434--3,435 & Flux increase in \fermi channels & $1.3\sigma$ at ${t = 3,\hspace{-1pt}434}$, and $1.7\sigma$ at  ${t = 3,\hspace{-1pt}435}$ & Detected at ${t = 3,\hspace{-1pt}434}$ and reported in ATel at ${t = 3,\hspace{-1pt}435}$ \citep[][]{fermi_atel_bllac_2019} \\

    3,436 & VHE flaring activity & $5\sigma$ at ${t = 3,\hspace{-1pt}436}$  & Detected by VERITAS and MAGIC at ${t = 3,\hspace{-1pt}436}$ and reported at ${t = 3,\hspace{-1pt}436}$ \citep[][]{magic_atel_bllac_2019}{}{}  \\

    3,470--3,500 & Return to low-state in VHE & $4\sigma$ at ${t = 3,\hspace{-1pt}467}$, and $2.7\sigma$ at ${t = 3,\hspace{-1pt}498}$ & No response   \\

    3,880--3,900 & Brightness increase in optical and gamma-ray flare & $3\sigma$ at ${t = 3,\hspace{-1pt}896}$, and $3.8\sigma$ at ${t = 3,\hspace{-1pt}899}$  &  Detected by Crimean Astrophysical Observatory and \fermil at ${t = 3,\hspace{-1pt}900}$ and reported at ${t = 3,\hspace{-1pt}901}$ \citep[][]{crim_telescope_atel_bllac_august_2020,fermi_atel_bllac_august_2020}  \\

    3,909--3,911 & Increase in optical and gamma-ray variability & $3\sigma$ at ${t = 3,\hspace{-1pt}909}$ & Detected at ${t = 3,\hspace{-1pt}910}$ by \fermil and MAGIC, reported at ${t = 3,\hspace{-1pt}913}$ \citep[][]{fermi_atel_bllac_august_2020_number_2, magic_atel_bllac_august_2020}. Optical brightening detected by Automatic Telescope for Optical Monitoring (ATOM) between ${t = 3,\hspace{-1pt}909}$ and ${t = 3,\hspace{-1pt}911}$, reported at ${t = 3,\hspace{-1pt}911}$, detected by Hans-Haffner-Sternwarte at ${t = 3,\hspace{-1pt}910}$, reported at ${t = 3,\hspace{-1pt}912}$ \citep[][]{atom_atel_bllac_august_2020, hans_haffner_atel_bllac_august_2020} \\

    3,940--3,941 & Gamma-ray flare & $2.4\sigma$ at ${t = 3,\hspace{-1pt}940}$ (driven by \fermi data) & Detected by MAGIC at ${t = 3,\hspace{-1pt}941}$ \citep[][]{magic_atel_bllac_september_2020}{}{}  \\

    3,952--3,957 & Increased gamma-ray activity &  $3.6\sigma$ at ${t = 3,\hspace{-1pt}951}$, and $5\sigma$ at ${t = 3,\hspace{-1pt}957}$ & Detected by \fermil collaboration from ${t = 3,\hspace{-1pt}957}$ onwards, reported ${t = 3,\hspace{-1pt}959}$ \citep[][]{fermi_atel_bllac_october_7_2020}. \\

    3,958 & Record X-ray flare & $10\sigma$ at ${t = 3,\hspace{-1pt}958}$ &  X-ray and UV rise detected by \textit{Swift}  at ${t = 3,\hspace{-1pt}957}$, reported at ${t = 3,\hspace{-1pt}958}$ \citep[][]{swift_atel_bllac_october_6_2020}, X-ray flare detected by \textit{Swift}-XRT at ${t = 3,\hspace{-1pt}958}$, reported at ${t = 3,\hspace{-1pt}959}$ \citep[][]{swift_atel_bllac_october_7_2020}{}{}  \\

    \bottomrule
    \end{tabular}
    \caption{Summary of the community and framework response to historical \mwl flaring events. The parameter, $t$, denotes the time in days since MJD 55,170;}
    \label{tab:mwlresponses}
\end{table*}

The panel on the right shows a prolonged phase of increasing \mwl activity in 2020, culminating in a brief, very bright X-ray flare at $t = 3,\hspace{-1pt}958$. From $t \sim 3,\hspace{-1pt}880$ onwards, the optical brightness slowly starts increasing, with the significance beginning to rise roughly seven days later. The significance reaches $3\sigma$
at $t = 3,\hspace{-1pt}896$, and $3.8\sigma$
at $t = 3,\hspace{-1pt}899$. A day later, the Crimean Astrophysical Observatory observed BL Lac in its historical optical r-band maximum. The \fermil collaboration observed a gamma-ray flare at the same time, both reported in ATels the following day \citep[][]{crim_telescope_atel_bllac_august_2020, fermi_atel_bllac_august_2020}.

Up to this point, the significance is driven purely by the continuing increase in the optical channels, and the brightening gamma-ray flux is not significantly detected by \framework. However, after a short period of fluctuation in the g-band and \fermil channels, \framework yields another $3\sigma$ detection for  ${t \sim 3,\hspace{-1pt}909}$. In this case, it is driven by a continued increase in gamma-ray variability. Observations with \fermil and the VHE gamma-ray telescope, MAGIC, revealed a flare a day later ($t = 3,\hspace{-1pt}910$). This was subsequently reported in ATels \citep[][]{fermi_atel_bllac_august_2020_number_2, magic_atel_bllac_august_2020}. No contemporaneous observations of this MAGIC-detected VHE flare were taken with VERITAS.

Simultaneously, continued optical observations with multiple instruments also found BL Lac in a high state at $t = 3,\hspace{-1pt}909$ \citep[][]{atom_atel_bllac_august_2020, hans_haffner_atel_bllac_august_2020}.
A further increase in \fermil flux is detected around $t = 3,\hspace{-1pt}940$ at $2.4\sigma$. At $t = 3,\hspace{-1pt}941$ the MAGIC collaboration again detected a bright VHE gamma-ray flare \citep[][]{magic_atel_bllac_september_2020}{}{}. Here, the framework yields $2.4 \sigma$ one day earlier.
At $t = 3,\hspace{-1pt}951$, the continued increase in gamma-ray variability is detected by \framework at $3.6\sigma$.

These increases in \mwl activity were used to trigger observations with the \textit{Swift} satellite for four days (3,954--3,957). The XRT observed the second-highest count rate from BL Lac up to that date, and the \textit{Swift} UV-Optical Telescope (UVOT) found the source to be in an elevated UV-optical state at the same time \citep[][]{swift_atel_bllac_october_6_2020}{}{}.
Further \textit{Swift} observations at $t = 3,\hspace{-1pt}958$ subsequently found a "record
X-ray flare", with the highest observed count rate from this source
\citep[][]{swift_atel_bllac_october_7_2020}{}{}. Also at $t = 3,\hspace{-1pt}958$,
\fermil observed a gamma-ray flare from BL Lac
\citep[][]{fermi_atel_bllac_october_7_2020}{}{}. Meanwhile, \framework reaches a
significance around $5\sigma$ a day earlier, at $t = 3,\hspace{-1pt}957$, driven by the high
flux in all \fermi channels, particularly the high-energy ones.
This significance then quickly exceeds $10\sigma$ with the detection of the exceptionally bright X-ray state at $t = 3,\hspace{-1pt}958$.
Such a high X-ray state is quite divergent from the data used to train \framework.
Accordingly, the reconstruction significance contributes strongly to the combined signal.

\subsection{Conclusions}

We demonstrate the ability of our model to reliably detect flares of the blazar, BL Lacertae,
as well as hint at their precursors.
Detected events can be further characterized with our model. This is accomplished by retracing the contributions of the different channels (or combinations thereof) to the final significance.
We show that the uncertainties on the \mwl observables can be propagated onto the corresponding significance for detection; this improves the reliability predictions.

Our method is suited for
real-time analyses.
Comparing to the historical response of the community, we show that \framework detects flaring states at the same time or earlier than the state-of-the-art.
The significance obtained from our model offers a simple, single-scalar metric to quantify anomalous states (and possible precursors) in \mwl light curves. Such a metric can be helpful for communication.
Our approach thus has the potential to improve the reaction of the community to
interesting events.

\section{Summary and Discussion}\label{sec:discussion}
In this work, we showcase a novel deep learning analysis framework, capable of detecting various types of anomalies in real-world, \mwl light curves. While handling varying cadences and gaps in observational coverage, our model differentiates source variability from noisy background activity.
As we avoid the need for a labeled training dataset of flaring states, we open the door to discovery of the unexpected.

We evaluate our framework on simulations, and on historical blazar light curves. We demonstrate that we can effectively detect flaring behavior consisting of clear high or low states, as well as of correlated signals across bands. An advantage of our approach is the standardized quantification of anomalous states in terms
of their significance. This can be used to robustly make consistent decisions on resource management for future observations. It also facilitates communication within the \mwl community,
where different definitions of a flare are commonly used.

\sstitlenoskip{Performance on simulations and on real data}
Using light curve simulations, we show that upward and downward fluctuations from the baseline state of a source are detectable. The associated significance scales with the intensity of deviations.
Strong fluctuations from the baseline can already be identified from a single data point. Fainter deviations or
those in noisy channels require multiple subsequent data points for meaningful detection, as one would expect. 

We evaluate real observations of blazar variability spanning a wide range of the electromagnetic spectrum. Our framework can reliably detect known historical flares as they occur in different channels. The respective contributions
to the detection significance
from different channels or combinations thereof
can also be traced.
Comparing our results to the reaction of the international community, we confirm that our framework can consistently pick up flares 
on the same timescales. In some cases, our model would have
provided hints of upcoming interesting activity earlier
than was achieved in practice. 

\sstitlenoskip{Timescales of source variability}
A crucial characteristic of blazars is their pronounced variability on various timescales, ranging from minutes to years. 
By design, an instance of \framework is  sensitive to variability on a specific  timescale. (This is adjustable per channel, given the minimum time-bin width parameter, \gammatimeaggr).
In the current study, we illustrate the sensitivity of our framework for variability timescales of days. Other timescales could be considered in the future,
\eg using an ensemble of trained pipelines,
covering timescales from hours to weeks. 

The choice of a timescale for the search window is informed by the science questions being investigated. It is also constrained by 
the availability of relevant  data. To resolve this, one could \eg augment existing data with simulations. These may span different quiescent
configurations, usable as new training samples. Such an approach also opens the door to tuning the sensitivity of the model to particular topologies of flares.

\sstitlenoskip{Role of data cleaning}
As \framework is trained on non-anomalous data, which rarely exists in practice, prior data cleaning is needed.
This may introduce biases, as the definition of anomalous source behavior is typically influenced by the expectations and goals of the analysis. 
If the cleaning procedure is too lenient, \framework will not be sensitive to subtle anomalies. Conversely, cleaning that is too strict leads to an increase in the number of false-positive signals. 
For some highly variable sources, identifying and excluding anomalous states can be challenging, especially when there are large gaps in coverage. To address this challenge,
our approach incorporates a statistical data cleaning technique based on the $\gammasignoise$ parameter.

\sstitlenoskip{Impact of model dimensionality}
In general, simple models may not capture the richness of a dataset, and could result in underfitting.
Increasing the number of dimensions enables \framework to model more complex source variability. However, this comes at the cost of additional computational complexity, and may result in overfitting. 

As part of the forecasting stage, we address possible underfitting by progressively increasing the size of the model.
This is moderated by verifying that the 
training score no longer improves past a certain
level of complexity. To avoid overfitting, we use a wide range of augmented and randomized realizations of the training dataset, as described in \autoref{subsec:training}.
We also incorporate standard regularization techniques.

Another crucial aspect of
\framework is
the complexity of the autoencoder, and in particular the size of the embedding vector. The number of components of the BGMM is also of great importance.
As described above, we gradually increase the complexity until the mapping
to clusters in the embedding space stops improving. We use a diverse collection of augmented and randomized inputs, as for the forecasting stage. Augmentation in this case also includes addition of upward and downward fluctuations to the data. Such examples reduce the reconstruction
error of the autoencoder for
potential flares, which are not present in the background training sample. The performance of the
autoencoder is also verified as part of inference, by inspection of the reconstruction error via \sigmarec. Systematically high values of the reconstruction error would indicate underfitting.

\sstitlenoskip{Resampling uncertainties on observables}
We have illustrated how our model lends itself to propagating uncertainties on input data. In particular, outliers with substantial uncertainties yield significance values with respectively large uncertainties. This allows one to effectively avoid false-positive predictions that are
related to low-quality data. 

In order to propagate uncertainties from observables to the corresponding significance, the input light curves are resampled. For the current study, the underlying assumption is that the provided uncertainties on inputs are statistically independent; more complicated correlations may be considered as needed. We also note that after resampling, individual values may be nonphysical (\eg correspond to negative fluxes). While a restriction on the resampling process could circumvent this, that would lead to an underestimation of the corresponding uncertainty.  In any case, the data-driven neural network approach is agnostic to 
such transformations on input data.

\sstitlenoskip{Impact of varying observation cadence}
We designed our framework to be robust against potentially confounding changes in cadence. Our solution, based on 
temporal weights, enhances the sensitivity for recent signals in sparse channels. 
The downside of this approach is that it does not
differentiate between recent low activity and lack of data. This may limit the capability of the model to accurately capture intricate temporal patterns within and across channels. More sophisticated weighting schemes may be considered to address specific use cases.

\sstitlenoskip{Future work and extensions}
Our approach can be utilized to detect and analyze transient phenomena in general; the
main constraint is that the signal is detectable via deviations from an established baseline
of background activity.
Our framework therefore sets the groundwork for a broad range of future work. 
In particular,
the pipeline is suitable for deployment as part of a real-time automated broker system, such as AMPEL \citep{nordin_2019_ampel}. It can therefore be used to quickly detect interesting activity and alert the community. 

Future work may \eg explore more sophisticated forecasting models. In particular, probabilistic modeling of the differences between predictions and real data could increase the robustness of the model. 

Furthermore, the framework could be extended to differentiate between known categories of activity, while remaining sensitive to unexpected anomalies. For instance,
separate significance metrics could be derived for specific types of variability, \eg
based on clustering in the embedding space.
This also has potential as a tool for comparing different theoretical emission models, or for making predictions and detecting
flares in the context of specific models.

\acknowledgments 
We would like to thank the following
people for numerous useful discussions:
Q.~Feng, 
M.~Gurwell,
C. McGrath,
M.~Negro,
D.~Parsons,
and
B.~Rani.
We would also like to thank the CTAO consortium and the \fermil and VERITAS collaborations for conducting internal courtesy reviews, and for providing useful feedback on this work.

This work is supported by the Helmholtz Einstein International 
Berlin Research School in Data Science (HEIBRiDS). 
This research was supported by the Helmholtz Weizmann
Research School on Multimessenger Astronomy, funded
through the Initiative and Networking Fund of the Helmholtz
Association, DESY, the Weizmann Institute, the Humboldt
University of Berlin, and the University of Potsdam.

This work made use of data supplied by the UK Swift Science Data Centre at the University of Leicester.

This research made use of the NASA/IPAC Infrared Science Archive, which is funded by the National Aeronautics and Space Administration and operated by the California Institute of Technology.

This research made use of observations from the Submillimeter Array, a joint project between the Smithsonian Astrophysical Observatory and the Academia Sinica Institute of Astronomy and Astrophysics, funded by the Smithsonian Institution and the Academia Sinica.
We recognize that Maunakea is a culturally important site for the indigenous Hawaiian people; we are privileged to study the cosmos from its summit.

This research has made use of the VizieR catalogue access tool, CDS,
Strasbourg, France \citep{10.26093/cds/vizier}. The original description 
of the VizieR service was published in \citet{vizier2000}.

This research made use of observations obtained with the Samuel Oschin Telescope 48-inch and the 60-inch Telescope at the Palomar
Observatory as part of the Zwicky Transient Facility project. ZTF is supported by the National Science Foundation under Grants
No. AST-1440341 and AST-2034437 and a collaboration including current partners Caltech, IPAC, the Oskar Klein Center at
Stockholm University, the University of Maryland, University of California, Berkeley, the University of Wisconsin at Milwaukee,
University of Warwick, Ruhr University, Cornell University, Northwestern University and Drexel University. Operations are
conducted by COO, IPAC, and UW.

The \textit{Fermi}-LAT Collaboration acknowledges generous ongoing support
from a number of agencies and institutes that have supported both the
development and the operation of the LAT as well as scientific data analysis.
These include the National Aeronautics and Space Administration and the
Department of Energy in the United States, the Commissariat \`a l'Energie Atomique
and the Centre National de la Recherche Scientifique / Institut National de Physique
Nucl\'eaire et de Physique des Particules in France, the Agenzia Spaziale Italiana
and the Istituto Nazionale di Fisica Nucleare in Italy, the Ministry of Education,
Culture, Sports, Science and Technology (MEXT), High Energy Accelerator Research
Organization (KEK) and Japan Aerospace Exploration Agency (JAXA) in Japan, and
the K.~A.~Wallenberg Foundation, the Swedish Research Council and the
Swedish National Space Board in Sweden. Additional support for science analysis during the operations phase is gratefully acknowledged from the Istituto Nazionale di Astrofisica in Italy and the Centre National d'\'Etudes Spatiales in France. This work performed in part under DOE Contract DE-AC02-76SF00515.

The authors thank the VERITAS collaboration for making available the data used in this publication. 

This research made use of the CTAO instrument
response functions, see \url{https://www.cta-observatory.org/science/cta-performance/} for more details.

\section*{Data Availability}

The code used in this work is available on GitLab under a
MIT license: \href{https://gitlab.desy.de/trans_finder/blazar_flares}{https://gitlab.desy.de/trans\_finder/blazar\_flares}.
The simulation (HDF5) data for this study have been deposited
on Zenodo under a Creative Commons Attribution (v4) license:
doi:\href{https://doi.org/10.5281/zenodo.14698916}{10.5281/zenodo.14698916}.

\facilities{\cta, \textit{Fermi}-LAT, SMA, \textit{Swift}-XRT, Tuorla, PTF, ZTF, IRSA, VERITAS} 

\clearpage
\newpage
\appendix

\section{Light curve simulations}\label{App:SIM_DATA}
This section describes in more detail the specifics of the light curve simulations of 1ES$~$1215$+$303.
Three representative examples of simulated \mwl light curves are shown in \autoref{fig:simulation-lc_panels}.

\subsection{\fermil}
The Fermitools\footnote{\url{https://github.com/fermi-lat/Fermitools-conda/}} package (version 2.2.0), a software package provided by the \fermil collaboration to facilitate analysis of telescope data, includes the \textit{gtobssim} tool for simulating LAT observations.
Inputs to these simulations are the instrument response functions (\irfs) to be used; the time range of the observations; a list of sources to be simulated; and parameters corresponding to each source, such as their positions in the sky; their average fluxes; and their spectral models.
As the goal is to train the model on pure background data, we simulate the source at a constant flux level, only perturbed by Poissonian- and instrument-induced noise. Because we are simulating the source in two adjacent energy bands, \fermil and \cta respectively, we make sure that the simulated spectra in both bands are connected. In order to achieve this, we use the \textit{SpectralTransient} class from \textit{gtobssim}, allowing us to simulate a log-parabolic spectrum.
As input to the simulation, we use the fluxes obtained for the 2017 non-flaring period defined by \citet{Valverde_2020} (Table 9).
We simulate a total of 1,000 consecutive days of \fermil observations between MJD 55,999 and 56,999. In our case, we use the latest LAT \irf (P8R3\_SOURCE\_V3), and choose to limit the simulation to the source in question. We use the latest templates for diffuse and galactic background components (gll\_iem\_v07.fits, iso\_P8R3\_SOURCE\_V3\_v1.txt). Additionally, we use the relevant spacecraft file for the time range of the simulations, obtained from \url{https://fermi.gsfc.nasa.gov/cgi-bin/ssc/LAT/LATDataQuery.cgi}.

After generating the simulated event files, we use \textit{fermipy} \citep[v. 1.2,][]{wood_2017}{}{} to perform a \fermil maximum-likelihood analysis, fitting the target source and the background components in three adjoining energy bins 
(100 -- 669~MeV, 
669~MeV -- 4.48~GeV, and 
4.48 -- 300~GeV)
over the full 1,000 days. 
We use a $15^\circ$ circular 
region of interest
(ROI) and apply standard event-quality cuts, \textit{(DATA\_QUAL$>$0)\&\&(LAT\_CONFIG==1)}, \textit{evclass=128, evtype=3}, \textit{zmax=90}.

Finally, a daily-binned light curve is produced by fitting the source parameters together with the normalization of the isotropic diffuse background for each one-day interval, within each energy bin. For all input light curves, low-confidence data are discarded beforehand. This corresponds to all data with a negative detection test statistic, and all data where the relative flux uncertainty, $\Delta \mathrm{Flux} / \mathrm{Flux}$, is larger than 5.

\subsection{\cta}

Simultaneously, we simulate light curves of 1ES$~$1215$+$303 as it would be observed by the upcoming \cta, using the \textit{gammapy} package \citep[v. 0.20.1][]{gammapy:2017}{}{} together with the public preliminary \cta \irf\footnote{\url{https://www.cta-
observatory.org/science/cta-performance/}}. Specifically, the \irf we use is \textit{Prod5-North-20deg-AverageAz-4LSTs09MSTs.18000s-v0.1.fits.gz}.
In order to simulate a \cta light curve, one has to supply the following inputs: the source coordinates; an assumed spectral model, including average flux and a spectral index; a time profile; and times and exposure durations for the individual observation runs. We use the \textit{ConstantTemporalModel} class to simulate a realistic IACT observing cadence, given actual observation times from the VERITAS dataset of the source in question. This allows us to keep a realistic mix of small and large gaps between individual observations.

As a proxy for the source state, we use the flux and spectral parameters of the 2017 quiescent state, as given in \citet{Valverde_2020}{}{}. The spectral model the simulations are based upon is the same log-parabola model as used for the \fermil simulations. We add to this a term accounting for EBL extinction \citep[][]{finke_2010}{}{}, which results in a sharper drop-off of the flux at VHE energies. The subsequent analysis of the simulated data is also performed in \textit{gammapy}. We 
divide the simulated data in multiple adjacent energy bins
(31.6 -- 79.4~GeV, 
79.4 -- 199.5~GeV, 
199.5~GeV -- 1.2598~TeV, and
1.2598 -- 12.589~TeV). 
We discard data with a relative flux uncertainty higher than 3.

\section{Multiwavelength data}\label{App:MWL_DATA}

\subsection{VERITAS}

VERITAS is an array of four imaging atmospheric Cherenkov telescopes, located at the Fred Lawrence Whipple Observatory outside Tuscon, Arizona (31.7$^\circ$ N, 110.0$^\circ$ W). The VERITAS instrument and performance are described in detail in~\cite{Holder2008, VERITASinstrument}. Of particular relevance for this study is the VERITAS sensitive energy range, which extends from $\sim$100 GeV to 30 TeV. VERITAS has observed BL Lacertae since 2008. We obtain a long-term VERITAS light curve of the source, spanning the years 2008--2020 and including 80 hours of quality-selected data (live time). Data were processed via the standard VERITAS calibration and reconstruction pipelines~\citep{Eventdisplay}. Data with relative uncertainty higher than 3 were discarded.

As mentioned above, we use the reconstructed flux of the source as input for our pipeline. The full dataset for BL Lac has not yet been published by VERITAS.
We therefore show the corresponding excess photon count rate, slightly modified by random noise in \autorefs{fig:blac-lc-and-results} and \ref{fig:result-highlights}. This count rate traces the relevant changes to the measured flux of the source, and is thus shown here as a proxy for the flux of the source.

\subsection{\fermil}

The \fermil data are retrieved from the public LAT database. In order to create a long-term light curve, we analyze the LAT data in the same three energy bins as for the simulation study, using identical event cuts. For each energy bin, we perform an ROI fit of all photons  within a $15^\circ$ circle centered on BL Lac. We take into account all sources from 4FGL-DR3 \citep[][]{4fgl_dr3} within a $20^\circ$ circle. This ensures that photons leaking from these sources into the ROI are adequately modelled. We produce a daily-binned light curve for each energy bin over the full duration, including all data with non-negative test statistic values. Data with relative uncertainty higher than 3 are discarded.

\subsection{\textit{Swift}-XRT}
The XRT on-board the \textit{Swift} satellite \citep[][]{gehrels_2004, burrows_2005}{}{} observes the X-ray sky at energies between $0.2$ and $10~$keV. We select all observations of BL Lac in the relevant time range and use the \textit{Swift} online analysis tool at \url{https://www.swift.ac.uk/user_objects/} to automatically create a count-rate light curve \citep[][]{evans_2007_swift}.
In order to have some information about the spectral evolution of the source, we make use of the fact that the provided light curve can be separated into two energy bins; the soft band, from 0.3 to 1.5 keV, and the hard band, from 1.5 to 10 keV. In each bin, we filter data having a relative uncertainty $\geq0.25$. We additionally filter data that could be affected by pile-up due to high count rates, particularly in the photon counting mode \citep[$>0.5~$counts/s; see \eg][]{ballet_1999}{}{}.

\subsection{Optical datasets}

\subsubsection{Tuorla}

The Tuorla observatory, with its 1~m optical telescope located in Finland, provides r-band magnitudes for a collection of TeV-blazars \citep[][]{nilsson_2018}{}{}  between 2002 and 2012. From this database, we obtain the r-band light curve of the blazar BL Lacertae for the relevant time range.

\subsubsection{PTF \& ZTF}

The Palomar Transient Factory was an optical monitoring program, using a collection of telescopes located in San Diego County, California between 2009 and 2012 \citep[][]{law_2009}{}{}. In 2017, this was replaced by ZTF, the Zwicky Transient Facility \citep[][]{bellm_2019a, bellm_2019b}{}{}. Both programs are focused on observing transients, but due to their monitoring method also provide long-term light curves of persistent sources, such as blazars. Datasets for both programs are publicly available. We make use of their BL Lacertae r- and g-band light curves, provided through the IRSA time-series tool \citep[][]{irsa_time_series_tool}{}{}. PTF and ZTF datasets have to be retrieved individually, and individual cleaning steps have to be taken. For PTF, we include data with \textit{goodflag == 1}, while for ZTF we filter data with \textit{catflag $\geq$ 32768}.

\subsubsection{Combining the datasets}

We join the three optical datasets into two light curves, one for the r-band and one for the g-band. 
ZTF supersedes PTF and only began operations five years after the end of the Tuorla program. All optical data after MJD 59,784, following a long gap, are therefore exclusively ZTF data. 

Some temporal overlap exists between the PTF and Tuorla datasets. Even though the photometry between the two telescopes differs, quasi-simultaneous observations from the instruments provide consistent magnitudes. We therefore combine the respective r-band data.

ZTF and PTF also use different, but very similar filters and photometric corrections. In the case of PTF, SDSS is used as a reference, while the Pan-STARRS1 (PS1) Survey is used for ZTF. Given the large gap between the latest PTF data and the first ZTF data, even a slight offset between calibrations would not impact our results.

As some PTF data are associated with very large uncertainties, we exclude all PTF data that have an absolute uncertainty larger than the largest uncertainty in the Tuorla and ZTF datasets. This leaves no g-band data before the start of operations of ZTF.

\subsection{SMA}

The 0.87$~$mm and 1.3$~$mm flux density data were obtained at the Submillimeter Array (SMA) near the summit of Mauna Kea (Hawaii). They were kindly provided to us by M.~Gurwell.  We filter all data that have a relative measurement uncertainty $\geq0.1$. BL Lacertae is included in an ongoing monitoring program at the SMA. It aims to determine the fluxes of compact extragalactic radio sources, which can be used as calibrators at mm wavelengths \citep[][]{gurwell_2007}{}{}.  Observations of available potential calibrators are from time to time observed for three to five minutes. The measured source signal strength is calibrated against known standard objects, typically solar system objects (Titan, Uranus, Neptune, or Callisto). Data from this program are updated regularly, and are available at the SMA website \url{http://sma1.sma.hawaii.edu/callist/callist.html}.

\bibliography{main}{}
\bibliographystyle{aasjournal}

\end{document}